\documentclass[aps,prb,10pt,amsmath,amssymb,twocolumn,longbibliography,superscriptaddress]{revtex4-2}
\usepackage{float}
\usepackage[pdftex]{graphicx,xcolor}
\usepackage{dcolumn}
\usepackage{bm}
\usepackage{ascmac}
\usepackage{array}
\usepackage{siunitx}
\usepackage{physics}
\usepackage{hyperref}
\hypersetup{
     colorlinks=true,
     citecolor=blue,
     linkcolor=blue,
     }
\begin{document}

\title {Theory of fractional corner charges in cylindrical crystal shapes}

\author {Hidetoshi Wada}
\affiliation{
Department of Physics, Institute of Science Tokyo, 2-12-1 Ookayama, Meguro-ku, Tokyo 152-8551, Japan}

\affiliation{
Department of Physics, Tokyo Institute of Technology, 2-12-1 Ookayama, Meguro-ku, Tokyo 152-8551, Japan\\
}

\author {Tiantian Zhang}
\affiliation{Institute of Theoretical Physics, Chinese Academy of Sciences, Beijing 100190, China\\
}

\author {Shuichi Murakami}
\affiliation{
Department of Physics, Institute of Science Tokyo, 2-12-1 Ookayama, Meguro-ku, Tokyo 152-8551, Japan}

\affiliation{
Department of Physics, Tokyo Institute of Technology, 2-12-1 Ookayama, Meguro-ku, Tokyo 152-8551, Japan\\
}

\affiliation{
International Institute for Sustainability with Knotted Chiral Meta Matter (WPI-SKCM2), Hiroshima University, Higashi-hiroshima, Hiroshima 739-0046, Japan}

\date{\today}

\begin{abstract}
     Recent studies showed that topologically trivial insulators may have fractionally quantized corner charges due to the topological invariant called a filling anomaly. 
     Such crystal shapes in three dimensions are restricted to vertex-transitive polyhedra, which are classified into spherical and cylindrical families. 
     The previous works derived formulas of the fractional corner charge for the spherical family, which corresponds to the tetrahedral and cubic space groups (SGs). 
     In this study, we derive all the corner charge formulas for the cylindrical family, which corresponds to the orthorhombic, tetragonal, hexagonal, and trigonal crystal shapes.
     We show that all the real-space formulas of the filling anomaly for the cylindrical SGs are universally determined by the total charges at the Wyckoff position (WP) $1a$. 
     Moreover, we derive the $k$-space formulas of the corner charge for the cylindrical cases with time-reversal symmetry (TRS). 
     From our results, we also show that CsLi$_{2}$Cl$_{3}$, KN$_{3}$, and Li$_{3}$N are candidate materials with a quantized corner charge by using the $ab$ $initio$ calculations. 
     Together with our previous work, we exhaust corner charge formulas for all the SGs and crystal shapes having quantized corner charges. 
\end{abstract}

\maketitle

\section{Introduction}
The discovery of topological phases of matter has had a profound impact on condensed matter physics.
These phases are characterized by gapless excitations localized at their boundaries and are robust against continuous deformations \cite{PhysRevB.76.045302,PhysRevB.27.6083,PhysRevLett.98.106803,PhysRevLett.95.146802,PhysRevLett.95.226801,PhysRevB.74.195312,doi:10.1126/science.1148047,doi:10.1126/science.1133734,PhysRevB.78.045426,PhysRevLett.106.106802,PhysRevB.96.245115,PhysRevB.91.161105,PhysRevB.95.081107,PhysRevB.90.165114,PhysRevX.7.041069,PhysRevB.98.081110,PhysRevLett.119.246402,Schindler_2018,doi:10.1126/sciadv.aat0346,PhysRevResearch.2.012067,PhysRevLett.124.036803,PhysRevResearch.2.043131,PhysRevX.11.041064,PhysRevB.105.045126,PhysRevB.109.085114,PhysRevB.99.245151,PhysRevB.103.205123,PhysRevResearch.1.033074,PhysRevB.102.165120,Po_2017,SI_SA_Watanabe,Bradlyn_2017,PhysRevB.97.035139,Aroyo:xo5013}. 
Moreover, topological materials exhibiting those features are characterized in terms of the classification indices called topological invariants, and those phases include the Chern insulators \cite{PhysRevB.76.045302,PhysRevB.27.6083}, $\mathbb{Z}_{2}$ topological insulators \cite{PhysRevLett.98.106803,PhysRevLett.95.146802,PhysRevLett.95.226801,PhysRevB.74.195312,doi:10.1126/science.1148047,doi:10.1126/science.1133734}, topological crystalline insulators (TCIs) \cite{PhysRevLett.106.106802,PhysRevB.96.245115,PhysRevB.91.161105,PhysRevB.95.081107,PhysRevB.90.165114,PhysRevX.7.041069}, and higher-order topological insulators (HOTIs) \cite{PhysRevB.98.081110,PhysRevLett.119.246402,Schindler_2018,doi:10.1126/sciadv.aat0346,PhysRevResearch.2.012067,PhysRevLett.124.036803,PhysRevResearch.2.043131}. 
Today, the field of topological phases of condensed matter physics has been attracting much attention in various fields. 
Moreover,  topological quantum chemistry \cite{Bradlyn_2017,PhysRevB.97.035139,Aroyo:xo5013, MTQC} and the theory of symmetry-based indicators \cite{Po_2017, SI_SA_Watanabe} allow us to identify many topological materials \cite{catalogue0,catalogue1,catalogue2,catalogue3,catalogue4,catalogue6}.

In recent years, it has been also discovered that even topologically trivial insulators called atomic insulators (AIs) have interesting features characterized by some topological invariants. 
Electronic states in AIs consist of exponentially localized Wannier orbitals, and their energy spectra are completely  gapped in the bulk.
Meanwhile, it was revealed that some AIs, which we call obstructed atomic insulators (OAIs), have fractionally quantized charges on their corners \cite{PhysRevB.105.045126,PhysRevB.109.085114,PhysRevB.99.245151,PhysRevX.11.041064,PhysRevB.103.205123,PhysRevResearch.1.033074,PhysRevB.102.165120}. 
OAIs are characterized by charge imbalance between electrons and ions at each WP, and the corner charge in OAIs can be understood in terms of a filling anomaly \cite{PhysRevB.99.245151,PhysRevB.103.165109}. 
A filling anomaly is a topological invariant, which is defined as difference between the number of electrons required by crystallographic symmetries of the system and the number of electrons required for charge neutrality. 
Fractional corner charges are given by equal distribution of the filling anomaly to all the corners. 

In the previous works \cite{PhysRevB.105.045126,PhysRevB.109.085114}, we focused on the tetrahedral and cubic SGs, and derived general corner charge formulas for various crystal shapes corresponding to those SGs. 
Fractional quantization of the corner charge requires the crystal shapes to be vertex-transitive polyhedra, where all the corners are equivalent by symmetry \cite{bridges2002:320,ROBERTSON197079,polyhedra,https://doi.org/10.1112/jlms/s2-2.1.125,IsogonalPrismatoids,Leopold,Grunbaum}.
Vertex-transitive polyhedra are classified into two families, the spherical family and cylindrical family, and in the previous works \cite{PhysRevB.105.045126,PhysRevB.109.085114}, we focused on the spherical family having tetrahedral and cubic symmetries. 
Then for the vertex-transitive polyhedra corresponding to the each SG, we obtained the corner charge formulas in real space. 
Moreover, to obtain the corner charge formulas in $k$ space, we took the method of the elementary band representation (EBR) matrix \cite{PhysRevB.105.045126,PhysRevB.109.085114,PhysRevB.103.165109}. 

In this study, we focus on the remaining family, namely the cylindrical family of the vertex-transitive polyhedra to derive a full list of formulas for the fractional corner charges.  
They correspond to orthorhombic, tetragonal, hexagonal, and trigonal SGs.  
We derive real-space formulas of the corner charges for all the crystal shapes in the cylindrical family with quantized fractional corner charge, and find that they are universally given by the charge imbalance at the Wyckoff position $1a$ divided by the number of corners of the crystal shapes.
Moreover, we derive the $k$-space formulas of the corner charge for the above SGs in the same way as the previous work. 
We find that in most of the cases, the $k$-space corner charge formulas are given by the number of irreps for the occupied bands at high-symmetry points.
However, for the SGs No.~89, 97, and 177, the $\mathbb{Z}_{2}$ topological invariant $\xi$, as defined for SG No.~16 and described in references \cite{PhysRevB.109.085114,Ono=Shiozaki_top_inv}, must also be taken into account in the formula to fully determine the corner charges. 
To summarize, this study and the previous work \cite{PhysRevB.109.085114} constitute a full list of $k$-space formulas of the corner charge with TRS for all the SGs and all the corresponding crystal shapes with a quantized corner charge.

This paper is organized as follows.
In Sec.~\ref{Sec.2}, we briefly review the previous studies on formulas of quantized corner charges. 
In Sec.~\ref{Sec.3}, we construct real-space formulas of the corner charge for the cylindrical crystal shapes, and discuss ambiguities in hinge charges and corner charges. 
In Sec.~\ref{Sec.4}, we construct a $k$-space formula by means of the EBR matrix. 
We also discuss the candidate materials with a quantized corner charge. 
Our conclusions are given in Sec.~\ref{Sec:5}.

\section{Review of fractional corner charges in various spherical crystal shapes \label{Sec.2}}

\begin{figure}[t]
    \centering
    \includegraphics[scale = 0.6]{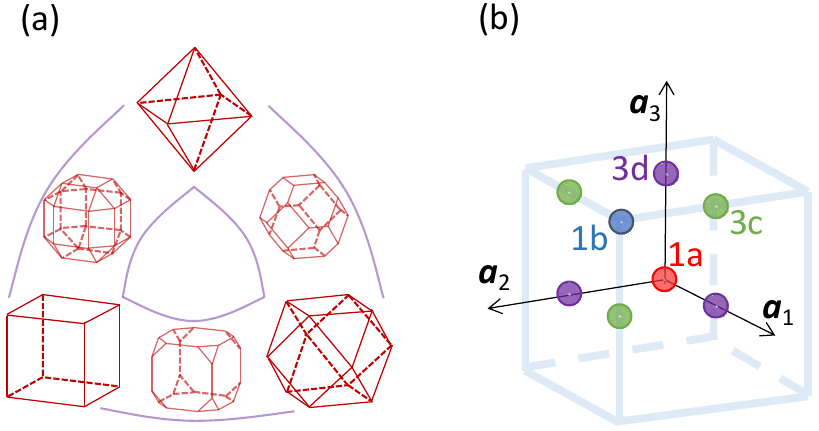}
    \caption{Calculation of the fractional corner charge in the SG No.~207 in the previous study \cite{PhysRevB.109.085114}. (a) Vertex-transitive polyhedra corresponding to the cubic point group $O$. In between the three polyhedra at the corners (cube, octahedron, and  cuboctahedron), there exist a continuous family of polyhedra. (b) Wyckoff positions (WPs) in the SG No.~207. The points with the same colors belong to the same WPs. $\bm{a}_{i}\ (i = 1, 2, 3)$ are the primitive lattice vectors in the cubic unit cell. The blue cube represents the unit cell.}
    \label{fig:207}
\end{figure}

\begin{figure*}[t]
    \centering
    \includegraphics[scale = 0.55]{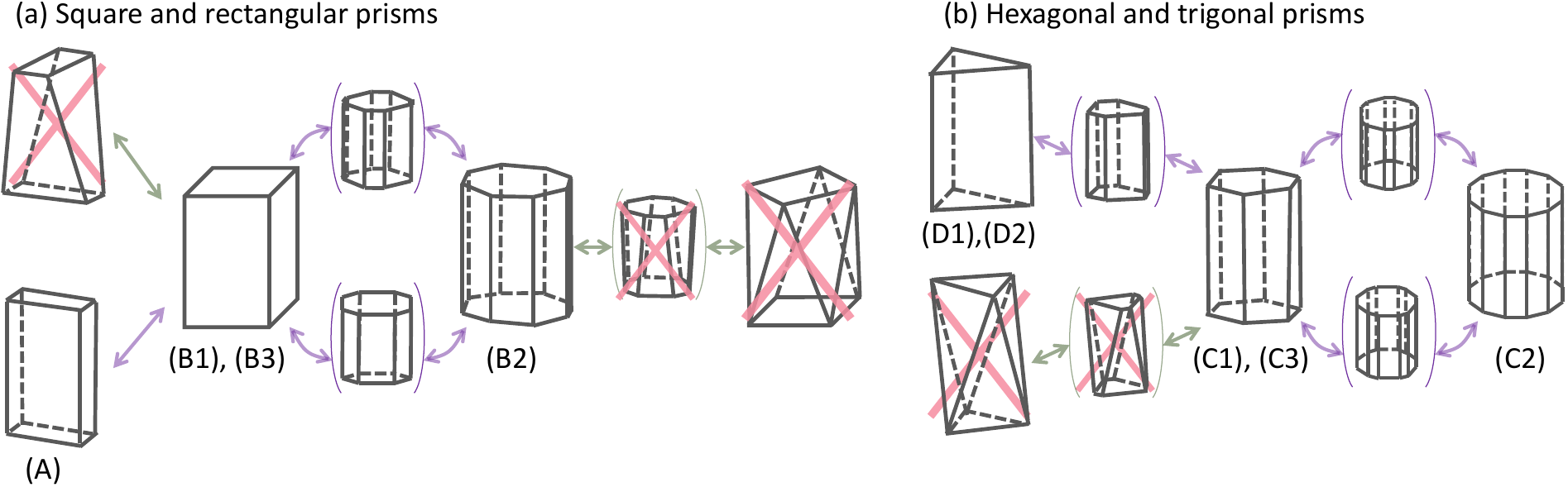}
    \caption{Vertex-transitive polyhedra in the cylindrical family \cite{IsogonalPrismatoids}. (a) Family of square and rectangular prisms. (b) Family of hexagonal and trigonal prisms. The green and purple arrows indicate the continuous deformation of the polyhedra preserving rotoreflection symmetry, and the cylindrical symmetry enumerated in Eq.~(\ref{Eq:point_group}), respectively. The polyhedra with the pink cross marks indicate that the corner charge is not quantized. The polyhedra denoted by (A), (B1)--(B3), (C1)--(C3) correspond to the crystal shapes with the same notation in Fig.~\ref{fig:crystal_shapes}.  }
    \label{Fig:vertex_prism}
\end{figure*}

In this section, we briefly review the concept of the fractional corner charges in three-dimensional crystals based on the previous works \cite{PhysRevB.105.045126,PhysRevB.109.085114}.
The theory of fractional corner charges in three-dimensional systems is an extension of Ref.~\cite{PhysRevB.99.245151}, which is first establish the theory of fractional corner charges in two dimensions.  
In the previous work \cite{PhysRevB.109.085114}, we extended the corner charge formulas for five crystal shapes with the SG No.~207 \cite{PhysRevB.105.045126} to all the crystal shapes in all the tetrahedral and cubic SGs. 
Quantization of the corner charge occurs only in vertex-transitive polyhedra, in which all the corners are related by symmetry operations.
In the tetrahedral and cubic SGs, such polyhedra are limited to the spherical family of the vertex-transitive polyhedra, whose classifications are given in Refs.~\cite{polyhedra,https://doi.org/10.1112/jlms/s2-2.1.125}. 
Here, we review the corner charge formula for the SG No.~207, which corresponds to the cubic point-group symmetry $O$. 
The vertex-transitive polyhedra under the point group (PG) $O$ are shown in Fig.~\ref{fig:207}(a).

\subsection{Setup}
In this subsection, we describe the setup required to derive the corner charge formulas. 
Firstly, we discuss properties of ionic and electronic charges in AIs. 
We consider ions to be made of nuclei and core electrons. 
Therefore, the ionic charges in AIs are integer multiples of the elementary charge $\abs{e}$, where $-e\ (e>0)$ is the electron charge. 
Moreover, electrons in AIs occupy Wannier orbitals of electronic bands, which are exponentially localized, and the integral charges of electrons in the Wannier orbitals can be assigned to the Wannier centers.
In the following, we assume that the gap is open both in the bulk and at the boundaries of AIs.  

We also review the filling anomaly \cite{PhysRevB.99.245151,PhysRevB.103.165109}. 
It is known that in certain insulators, the charge neutrality condition can be broken in some crystal shapes while the entire system is insulating and charge distribution preserves the crystallographic symmetry. 
In this case, the deviation from the charge neutrality condition is called the filling anomaly. 

Now, we consider the filling anomaly in terms of maximal WPs. 
Here, we focus on the cubic SG No.~207, with its unit cell shown in Fig.~\ref{fig:207}(b).
Let $n_{\omega}$ be the number of occupied Wannier orbitals of electrons at the maximal WP $\omega (= a, b, c, d)$ and $m_{\omega}$ be the total charge of ions measured in the unit of the elementary charge $\abs{e}$ at a maximal WP $\omega$. 
We then define $\Delta\omega$ to be the difference between them:
\begin{align}
    \Delta\omega = n_{\omega} - m_{\omega},
    \label{eq:delta}
\end{align}
where $\Delta\omega$ is always an integer by definition. 
We note that non-maximal WPs can be reduced to the maximal ones via continuous transformations, so we do not need to consider them. 

In order to calculate the filling anomaly, we need to specify the crystal shape.
For the quantization of the corner charge, we restrict the crystal shapes to be the vertex-transitive polyhedra with genus $0$ \cite{polyhedra,bridges2002:320,ROBERTSON197079,https://doi.org/10.1112/jlms/s2-2.1.125,IsogonalPrismatoids}, in which all of the corners are equivalent by the crystallographic symmetries. 
In the present case of the cubic PG $O$, the list of such polyhedra is shown in Fig.~\ref{fig:207}(a) \cite{polyhedra,https://doi.org/10.1112/jlms/s2-2.1.125}.
Then we can calculate the filling anomaly only from the values of $\Delta \omega$ for a specific crystal shape, such as a cube and an octahedron.
We assume the center of the crystal to be the maximal WP $1a$.
To obtain the filling anomaly, we first assume a perfect crystal in which the electronic states and the ionic positions are perfectly periodic in the whole crystal including its boundaries. 
Later, we will discuss how the things change if the electronic states and ionic positions near the boundaries are modulated. 
Then, the filling anomaly $\eta_{n}$ for a finite-sized perfect crystal with $n$ hinge periods can be expressed as a polynomial of $n$:
\begin{align}
    \eta_{n} = \alpha_{3} n^{3} + \alpha_{2} n^{2} + \alpha_{1} n + \alpha_{0}, 
    \label{eq:filling_anomaly}
\end{align}
where $\alpha_{i}\ (i = 0, 1, 2, 3)$ are integer constants, which depend on crystal shapes. 

Actually, $\alpha_{3}$ is proportional to the bulk charge density (per bulk unit cell), and $\alpha_{2}$ is proportional to the surface charge density (per surface unit cell) under the bulk charge neutrality condition $\alpha_{3} = 0$. 
Under the bulk and surface charge neutrality conditions $\alpha_{2} = \alpha_{3} = 0$, $\alpha_{1}$ is proportional to the hinge charge density (per hinge unit cell). 
Hereafter, we assume that the bulk, surfaces, and hinges are all charge neutral. 
In this case, the filling anomaly $\eta_{n}$ in Eq.~(\ref{eq:filling_anomaly}) is given by $\alpha_{0}$. 

\subsection{Real-space formulas of corner charges in all the possible crystal shapes with SG No.~207}
Because all the excess charge, i.e. the filling anomaly, should be at the corners, the corner charge is given by the filling anomaly divided by the number of corners \cite{PhysRevB.105.045126,PhysRevB.109.085114,PhysRevB.99.245151,PhysRevB.103.205123}, i.e.  
\begin{align}
    Q_{\text{corner}} = -\frac{\alpha_{0}}{N}\abs{e}\ \left( \text{mod} \frac{2\abs{e}}{N} \right),
\end{align}
where $N$ represents the number of corners of the crystal
shape. 
Notably, this formula holds even when electronic states and ionic positions near the boundaries of the crystal are modulated, since such modulations change the corner charge $Q_{\text{corner}}$ only by an integer multiple of $2\abs{e}/N$. 
Related discussions are given in Sec. \ref{Sec.3}. 

In the present case of the SG No.~207, after a lengthy calculation for all the possible crystal shapes with a quantized corner charge shown in Fig.~\ref{fig:207}(a), it was found that the filling anomaly is given by $\alpha_{0} = \Delta a\ (\equiv n_{a}-m_{a})$ (mod~2).
As a result, the real-space corner charge formulas for the various crystal shapes in the SG No.~207 are universally given by 
\begin{align}
    Q_{\text{corner}} = -\frac{\Delta a}{N}\abs{e}\ \left( \text{mod} \frac{2\abs{e}}{N} \right).
    \label{eq:corner_charge_207}
\end{align}  
Here the Wyckoff position $1a$ appears in the formula, because it is the WP located at the center of the crystal. 
We can find that this result is of the same form as that in two dimensions \cite{PhysRevB.103.205123,PhysRevB.102.165120}.
A similar formula was obtained for all the tetrahedral and cubic SGs and for all the crystal shapes with a quantized corner charge \cite{PhysRevB.109.085114}. 

\section{Corner charge formulas for various cylindrical crystal shapes \label{Sec.3}}

In Sec. \ref{Sec.2}, we reviewed the corner charge formula of the spherical family of vertex-transitive polyhedra. 
In fact, vertex-transitive polyhedra with genus $0$ are classified into spherical and cylindrical families \cite{polyhedra,https://doi.org/10.1112/jlms/s2-2.1.125,IsogonalPrismatoids}. 
In the present paper, we focus on the cylindrical family. 
There are many types of vertex-transitive polyhedra in the cylindrical family as shown in Fig.~\ref{Fig:vertex_prism}.
Moreover, in order to obtain the corner charge formulas, we find that it is enough to consider cuboids, tetragonal prisms, hexagonal prisms, and triangular prisms corresponding to the orthorhombic, tetragonal, hexagonal, and trigonal classes, respectively as we explain later. 
In this section, we calculate the corner charge formulas for those crystal shapes in terms of the bulk WPs. 
We find that the corner charges for the trigonal cases are always trivial, and the reason will be explained in Sec.~\ref{Sec.3}C. 

\subsection{Symmetries of various cylindrical crystal shapes} 
\begin{figure*}[t]
    \centering
    \includegraphics[scale = 0.6]{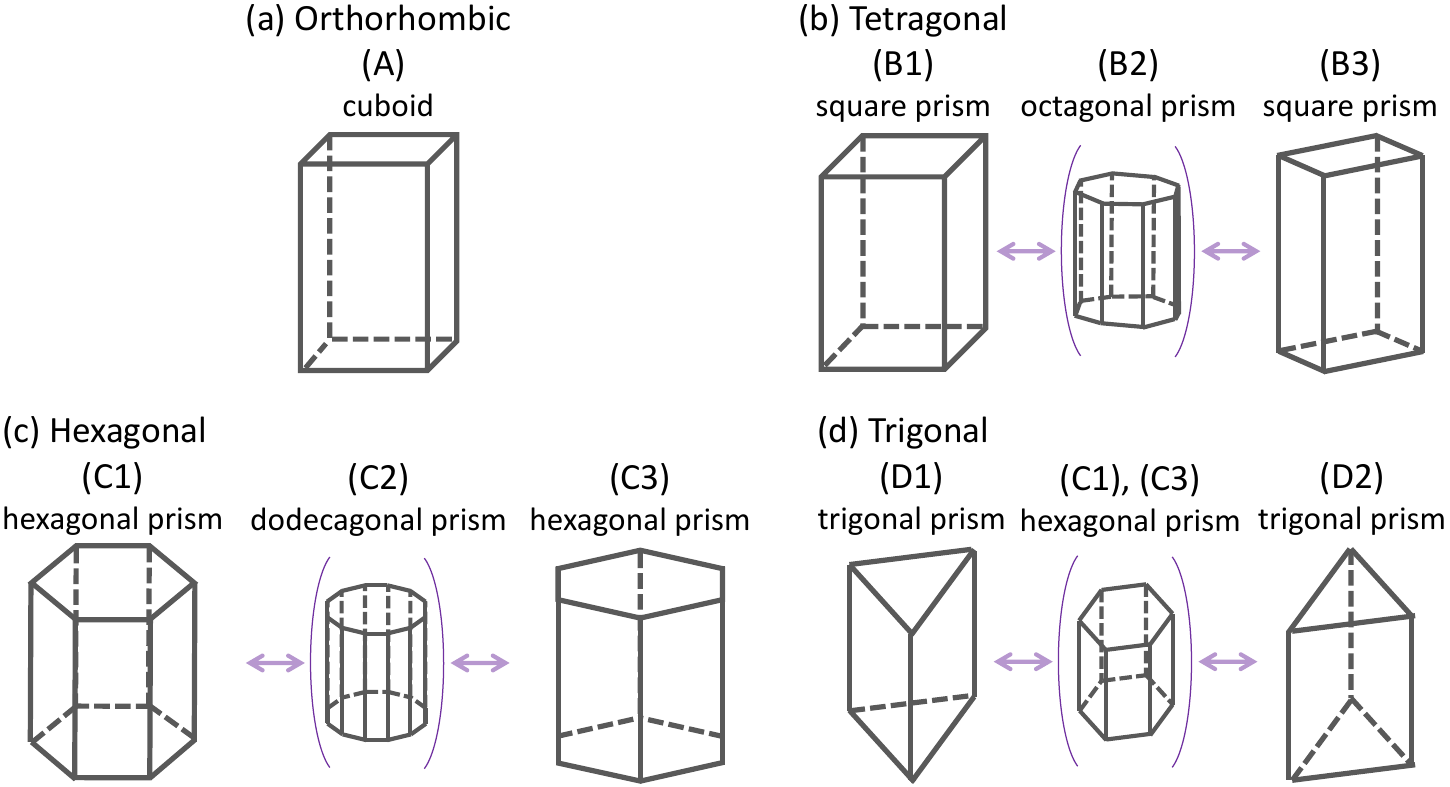}
    \caption{Summary of the crystal shapes with quantized corner charges in (a) orthorhombic, (b) tetragonal, (c) hexagonal, and (d) trigonal crystal shapes. The purple arrows in (b), (c) and (d) indicate continuous deformations of the crystal shapes. (B2) is in between (B1) and (B3), and thus its surface, hinge and corner charges can be derived from those for (B1) and (B3). Likewise, the surface, hinge and corner charges in (C2) can be derived from those for (C1) and (C3).}
    \label{fig:crystal_shapes}
\end{figure*}

The cylindrical family of vertex-transitive polyhedra consists of a wide variety of polyhedra, and in terms of the crystallographic PG, they are classified into prisms and antiprisms, whose bases are a square, a rectangle, a regular hexagon, and a regular triangle as shown in Fig.~\ref{Fig:vertex_prism} \cite{IsogonalPrismatoids}.  
We find that corner charges in the antiprisms i.e., the polyhedra with the pink cross marks in Fig.~\ref{Fig:vertex_prism} are not quantized, because these polyhedra contain surfaces with variable orientations \cite{PhysRevB.109.085114}. 
Namely, continuous changes of the surface orientations cause changes of the hinge charges, and therefore obstruct quantization of the hinge charges. 
Hence in these crystal shapes, corner charges are not quantized, and therefore we exclude them from our study. 
As a result, the vertex-transitive cylindrical crystal shapes exhibiting quantized corner charges are summarized in Fig.~\ref{fig:crystal_shapes}.

We also note that in order to obtain the corner charge formulas for all of the vertex-transitive polyhedra, we only have to consider only the following crystal shapes: cuboid in Fig.~\ref{fig:crystal_shapes}(A) for the orthorhombic crystal shapes, square prisms in Figs.~\ref{fig:crystal_shapes}(B1) and (B3) for the tetragonal crystal shapes, hexagonal prisms in Figs.~\ref{fig:crystal_shapes}(C1) and (C3) for the hexagonal crystal shapes, and trigonal prisms in Figs.~\ref{fig:crystal_shapes}(D1) and (D2) for the trigonal crystal shapes.
It is because the corner charge formulas for the other vertex-transitive polyhedra can be obtained from their results. 
For example, from Fig.~\ref{fig:crystal_shapes}(b), corner charge formulas for the octagonal prisms (Fig.~\ref{fig:crystal_shapes}(B2)) can be derived from those for the square prisms. 
In the same way, from Fig.~\ref{fig:crystal_shapes}(c), corner charge formulas for the dodecagonal prisms (Fig.~\ref{fig:crystal_shapes}(C2)) can be derived from those for the hexagonal prisms.
It is similar to the discussion in the previous works on the spherical family of polyhedra \cite{PhysRevB.105.045126,PhysRevB.109.085114}. 

As discussed in Sec.~\ref{Sec.2}, all the corners of these crystal shapes are connected by symmetry operations, which guarantees quantization of the corner charges. 
The corresponding point-group symmetries for each crystal shape are enumerated as follows: 
\begin{equation}
    \begin{aligned}[b]
        \text{orthorhombic}:& D_{2h},  \\
        \text{tetragonal}:& D_{4h}, D_{2h}, D_{4}, D_{2d}, C_{4h}, \\
        \text{hexagonal}:& D_{6h}, D_{3h}, D_{6}, C_{6h}, \\
        \text{trigonal}:& D_{3h}, D_{3}, D_{3d}, C_{3h}.
        \label{Eq:point_group}
    \end{aligned}
\end{equation}
We have to consider the SGs such that the point-group symmetries for the WPs at the center of the crystal are among the list in Eq.~(\ref{Eq:point_group}). 
These SGs corresponding to orthorhomic, tetragonal, and hexagonal crystal shapes are shown in Tabs.~\ref{tab:corner_charge_orth}, \ref{tab:corner_charge_tetra}, and \ref{tab:corner_charge_hexa}, respectively. 
The trigonal cases are omitted here, since we find that the corner charges are always trivial, as explained in Sec.~\ref{Sec.3}C. 
Here we note that some SGs have several different PGs in Eq.~(\ref{Eq:point_group}) depending on the WPs at the crystal center. 
For example, the SG No.~123 ($P$4/$mmm$) has the orthorhombic crystal shape of a cuboid as shown in Fig.~\ref{fig:crystal_shapes}(a) with its center at the WP~$e$ (see Tab.~\ref{tab:corner_charge_orth}), and the tetragonal crystal shape of a square prism as shown in Fig.~\ref{fig:crystal_shapes}(b) with its center at the WP~$a$ (see Tab.~\ref{tab:corner_charge_tetra}), and their corresponding PGs are different. 

\subsection{Derivation of corner charge formula for cylindrical crystal shapes}
In this subsection, we derive real-space formulas of corner charges. 
As described in Sec.~\ref{Sec.2}A, we first assume a perfect crystal, and count the number of electrons and ions at maximal WPs using Eq.~(\ref{eq:delta}) in the bulk unit cell. 
We also introduce charge neutrality conditions for the bulk, surfaces, and hinges, and derive the corner charge formulas for the various cylindrical crystal shapes. 

\begin{figure}[b]
    \centering
    \includegraphics[scale = 0.65]{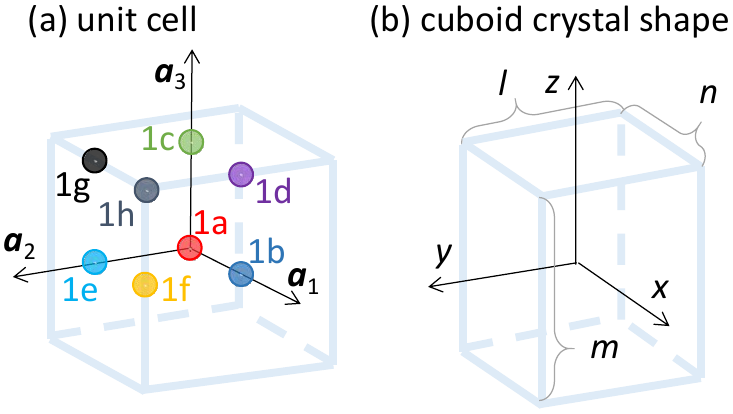}
    \caption{Calculation of the corner charge for the SG No.~47 ($Pmmm$). (a) Wyckoff positions (WPs) in the SG No.~47. $\bm{a}_{i}\ (i = 1, 2, 3)$ are primitive lattice vectors in the orthorhombic unit cell. The blue cuboid represents the unit cell. (b) Crystal shape of a cuboid with orthorhombic symmetry consisting of $l \times m \times n$ unit cells. $x$, $y$ and $z$-axes are parallel to $\bm{a}_{1}$, $\bm{a}_{2}$ and $\bm{a}_{3}$, respectively.}
    \label{fig:47}
\end{figure}

\begin{table}[h]
    \caption{Summary of WPs at the center of the crystal and their PGs for the orthorhombic and tetragonal SGs, and the real-space formulas of corner charges.}
    \centering
    \begin{ruledtabular}
    \begin{tabular}{cccc}
        SG number & Center WP & PG & $Q_{\text{corner}}(\text{mod}\ 2/N)$ \\
        \hline
        47 & $a$ & $D_{2h}$ & $\Delta a/N$ \\
        65 & $a$ & $D_{2h}$ & $\Delta a/N$ \\
        69 & $a$ & $D_{2h}$ & $\Delta a/N$ \\
        71 & $a$ & $D_{2h}$ & $\Delta a/N$ \\
        123 & $e$ & $D_{2h}$ & $\Delta e/N(\equiv \Delta a/N)$ \\
        127 & $c$ & $D_{2h}$ & $\Delta c/N(\equiv \Delta a/N)$ \\
        131 & $a$ & $D_{2h}$ & $\Delta a/N$ \\
        132 & $a$ & $D_{2h}$ & $\Delta a/N$ \\
        136 & $a$ & $D_{2h}$ & $\Delta a/N$ \\
        139 & $c$ & $D_{2h}$ & $\Delta c/N(\equiv \Delta a/N)$ \\
        140 & $d$ & $D_{2h}$ & $\Delta d/N(\equiv \Delta a/N)$ \\
    \end{tabular}
    \end{ruledtabular}
    \label{tab:corner_charge_orth}
\end{table}

\subsubsection{Orthorhombic}
Firstly, we explain the orthorhombic cases. 
The corresponding SGs are shown in Tab.~\ref{tab:corner_charge_orth}. 
Here, we take the SG No.~47 as an example, whose unit cell is shown in Fig.~\ref{fig:47}(a). 
We discuss a crystal shape of a cuboid consisting of $l \times m \times n$ unit cells as shown in Fig.~\ref{fig:47}(b). 
By using Eq.~(\ref{eq:delta}), the filling anomaly $\eta^{\text{orth}}_{l,m,n}$ is given by 
\begin{equation}
    \begin{aligned}[b]
        \eta^{\text{orth}}_{l,m,n} =& (\Delta a + \Delta b + \Delta c + \Delta d + \Delta e + \Delta f \\
        & + \Delta g + \Delta h)lmn \\
        &+ (\Delta a + \Delta c + \Delta e + \Delta g)lm \\
        &+ (\Delta a + \Delta b + \Delta c + \Delta d)mn \\
        &+ (\Delta a + \Delta b + \Delta e + \Delta f)ln \\
        &+ (\Delta a + \Delta e)l + (\Delta a + \Delta c)m \\
        &+ (\Delta a + \Delta b)n + \Delta a. \label{eq:filling_anomaly_orth}
    \end{aligned}
\end{equation}
Similar to Eq.~(\ref{eq:filling_anomaly}) in the cubic case, the filing anomaly $\eta^{\text{orth}}_{l,m,n}$ can be expressed as a polynomial of $l$, $m$, and $n$. 
Now we impose the charge neutrality conditions for the bulk, surfaces, and hinges. 
From Fig.~\ref{fig:47}(a), the bulk charge density $\rho_{\text{bulk}}$ (per bulk unit cell) is given by $\rho_{\text{bulk}}=-\abs{e}(\Delta a + \Delta b + \Delta c + \Delta d + \Delta e + \Delta f + \Delta g + \Delta h)$. 
Then the charge neutrality condition for the bulk is 
\begin{align}
    \Delta a + \Delta b + \Delta c + \Delta d + \Delta e + \Delta f + \Delta g + \Delta h = 0. 
    \label{eq:charge_neutral_bulk}
\end{align}
Therefore the first term in the right hand side in Eq.~(\ref{eq:filling_anomaly_orth}) can be dropped. 

Next let us investigate the surface charges. 
According to the modern theory of polarization \cite{PhysRevB.48.4442,PhysRevB.47.1651}, the bulk polarization is given by 
\begin{equation}
    \begin{aligned}[b]
        \bm{P}_{\text{bulk}} = \frac{-\abs{e}}{2v}&\bigl((\Delta b + \Delta d + \Delta f + \Delta h)\bm{a}_{x} \\
        & + (\Delta e + \Delta f + \Delta g + \Delta h)\bm{a}_{y} \\
        & + (\Delta c + \Delta d + \Delta g + \Delta h)\bm{a}_{z}\bigr)\ \ \left(\text{mod} \frac{\bm{R}}{v}|e|\right),
        \label{Eq:P_bulk_orth}
    \end{aligned}
\end{equation}
where $\bm{a}_{i}~(=a_{i}\bm{e}_{i})~(i=x,y,z)$ is the primitive lattice vector in the cubic unit cell, $a_{i}~(i=x,y,z)$ is the lattice constant, and $v~(=a_{x}a_{y}a_{z})$ is the volume of the unit cell. 
$\bm{R}=\Sigma_{i =x,y,z}m_{i}\bm{a}_{i}~(m_{i}\in\mathbb{Z})$ is a lattice vector. 
From Eq.~(\ref{Eq:P_bulk_orth}), the surface charge densities $\sigma^{i}_{\text{sur}}\ (\equiv\bm{P}_{\text{bulk}}\cdot\bm{e}_{i}a_{j}a_{k})~(i,j,k=x,y,z,\ i \neq j \neq k \neq i)$ per surface unit cell for the surface normal to the $i$ axis are expressed as 
\begin{equation}
    \begin{aligned}[b]
        \sigma^{x}_{\text{sur}} \equiv& \frac{-\abs{e}}{2}(\Delta b + \Delta d + \Delta f + \Delta h) \\
        \equiv& \frac{-\abs{e}}{2}(\Delta a + \Delta c + \Delta e + \Delta g)\ \ (\text{mod}\ \abs{e}),
    \end{aligned}
\end{equation}
\begin{equation}
    \begin{aligned}[b]
        \sigma^{y}_{\text{sur}} \equiv& \frac{-\abs{e}}{2}(\Delta e + \Delta f + \Delta g + \Delta h)\\ 
        \equiv& \frac{-\abs{e}}{2}(\Delta a + \Delta b + \Delta c + \Delta d)\ \ (\text{mod}\ \abs{e}),
    \end{aligned}
\end{equation}
\begin{equation}
    \begin{aligned}[b]
        \sigma^{z}_{\text{sur}} \equiv& \frac{-\abs{e}}{2}(\Delta c + \Delta d + \Delta g + \Delta h) \\
        \equiv& \frac{-\abs{e}}{2}(\Delta a + \Delta b + \Delta e + \Delta f)\ \ (\text{mod}\ \abs{e}),
    \end{aligned}
\end{equation}
which agree with the coefficients of the second-order terms in $l$, $m$ and $n$ in Eq.~(\ref{eq:filling_anomaly_orth}).
Then the charge neutrality conditions for the surfaces are 
\begin{align}
    \Delta a + \Delta c + \Delta e + \Delta g &\equiv 0\ \ (\text{mod}\ 2), \label{eq:charge_neutral_sur1} \\
    \Delta a + \Delta b + \Delta c + \Delta d &\equiv 0\ \ (\text{mod}\ 2), \label{eq:charge_neutral_sur2}\\
    \Delta a + \Delta b + \Delta e + \Delta f &\equiv 0\ \ (\text{mod}\ 2). \label{eq:charge_neutral_sur3}
\end{align}

Under these conditions, the second-order terms for $l$, $m$, and $n$ in Eq.~(\ref{eq:filling_anomaly_orth}) can also be dropped: 
\begin{equation}
    \begin{aligned}
        \eta^{\text{orth}}_{l,m,n} =& (\Delta a + \Delta e)l + (\Delta a + \Delta c)m \\
        &+ (\Delta a + \Delta b)n + \Delta a.
        \label{eq:filling_anomaly_orth2}
    \end{aligned}
\end{equation}
From this equation, we can calculate the hinge charge densities $\lambda_{\text{hinge}}^{l}$, $\lambda_{\text{hinge}}^{m}$ and $\lambda_{\text{hinge}}^{n}$ per unit cell along the hinge of length $l$, $m$ and $n$, respectively. 
Since there are four equivalent hinges of length $l$, $m$ and $n$ in the cuboid, the total charges of these hinges are $4i\lambda_{\text{hinge}}^{i}$, where $i=l$, $m$, or $n$.
Thus, from Eq.~(\ref{eq:filling_anomaly_orth2}), the hinge charge densities $\lambda_{\text{hinge}}^{i}~(i = l,m,n)$ are given by 
\begin{align}
    \label{eq:hinge_l_ortho}
    \lambda_{\text{hinge}}^{l} &= \frac{-\abs{e}}{4}(\Delta a + \Delta e), \\
    \lambda_{\text{hinge}}^{m} &= \frac{-\abs{e}}{4}(\Delta a + \Delta c), \\
    \lambda_{\text{hinge}}^{n} &= \frac{-\abs{e}}{4}(\Delta a + \Delta b),
\end{align}
all of which are defined in terms of mod $\abs{e}/2$ from Eqs~(\ref{eq:charge_neutral_bulk}) and (\ref{eq:charge_neutral_sur1})--(\ref{eq:charge_neutral_sur3}). 

Therefore, under the charge neutrality conditions for the hinges, we obtain the real-space formula of corner charges for the cuboid with the SG No. 47: 
\begin{align}
    Q_{\text{corner}} = -\frac{\Delta a}{8}|e|\ \ \left(\text{mod}\ \frac{\abs{e}}{4} \right). 
    \label{eq:corner_ortho}
\end{align}
Likewise, we can calculate the corner charge for all the orthorhombic SGs with quantized corner charges.
The summary of the corner charge formulas in the orthorhombic SGs are shown in Tab.~\ref{tab:corner_charge_orth}. 

\begin{table}[h]
    \caption{Summary of WPs at the center of the crystal and their PGs for the tetragonal SGs, and the real-space formulas of corner charges. There are some SGs, in which their corner charges are trivial. We note that some SGs have different PGs depending on the center WPs.}
    \centering
    \begin{ruledtabular}
    \begin{tabular}{cccc}
        SG number & Center WP & PG & $Q_{\text{corner}}(\text{mod}\ 2/N)$ \\
        \hline
        83 & $a$ & $C_{4h}$ & $\Delta a/N$ \\
        87 & $a$ & $C_{4h}$ & $\Delta a/N$ \\
        89 & $a$ & $D_{4}$ & $\Delta a/N$ \\
        97 & $a$ & $D_{4}$ & $\Delta a/N$ \\
        111 & $a$ & $D_{2d}$ & $\Delta a/N$ \\
        115 & $a$ & $D_{2d}$ & $0$ \\
        119 & $a$ & $D_{2d}$ & $0$ \\
        121 & $a$ & $D_{2d}$ & $\Delta a/N$ \\
        123 & $a$ & $D_{4h}$ & $\Delta a/N$ \\
        124 & $a$ & $D_{4}$ & $\Delta a/N$ \\
            & $b$ & $C_{4h}$ & $\Delta b/N(\equiv \Delta a/N)$ \\
        125 & $a$ & $D_{4}$ & $\Delta a/N$ \\
          & $c$ & $D_{2d}$ & $\Delta c/N(\equiv \Delta a/N)$ \\
        126 & $a$ & $D_{4}$ & $\Delta a/N$ \\
        127 & $a$ & $C_{4h}$ & $\Delta a/N$ \\
        128 & $a$ & $C_{4h}$ & $\Delta a/N$ \\
        129 & $a$ & $D_{2d}$ & $0$ \\
        131 & $e$ & $D_{2d}$ & $0$ \\
        132 & $b$ & $D_{2d}$ & $\Delta b/N(\equiv \Delta a/N)$ \\
        134 & $a$ & $D_{2d}$ & $\Delta a/N$ \\
        137 & $a$ & $D_{2d}$ & $0$ \\
        139 & $a$ & $D_{4h}$ & $\Delta a/N$ \\
            & $d$ & $D_{2d}$ & $0$\\
        140 & $a$ & $D_{4}$ & $\Delta a/N$ \\
            & $b$ & $D_{2d}$ & $\Delta b/N(\equiv \Delta a/N)$ \\
            & $c$ & $C_{4h}$ & $\Delta c/N(\equiv \Delta a/N)$ \\
        141 & $a$ & $D_{2d}$ & $0$
    \end{tabular}
    \end{ruledtabular}
    \label{tab:corner_charge_tetra}
\end{table}

\begin{table}[b]
    \caption{Summary of WPs at the center of the crystal and their PGs for the hexagonal SGs, and the real-space formulas of corner charges. There are some SGs, in which their corner charges are trivial. We note that some SGs have different PGs depending on the center WPs.}
    \centering
    \begin{ruledtabular}
    \begin{tabular}{cccc}
        SG number & Center WP & PG & $Q_{\text{corner}}(\text{mod}\ 2/N)$ \\
        \hline
        175 & $a$ & $C_{6h}$ & $\Delta a/N$ \\
        177 & $a$ & $D_{6}$ & $\Delta a/N$ \\
        187 & $a$ & $D_{3h}$ & $0$ \\
        189 & $a$ & $D_{3h}$ & $0$ \\
        191 & $a$ & $D_{6h}$ & $\Delta a/N$ \\
            & $c$ & $D_{3h}$ & $0$ \\
        192 & $a$ & $D_{6}$ & $\Delta a/N$ \\
            & $b$ & $C_{6h}$ & $\Delta b/N(\equiv \Delta a/N)$ \\
        193 & $a$ & $D_{3h}$ & $0$ \\
        194 & $b$ & $D_{3h}$ & $0$\\
    \end{tabular}
    \end{ruledtabular}
    \label{tab:corner_charge_hexa}
\end{table}

\subsubsection{Tetragonal}
As for the tetragonal crystal shapes, we derive the corner charge formulas in the same way as in the orthorhombic cases. 
In this case, it is enough to calculate the corner charge for the two tetragonal crystal shapes i.e. the square prism in Fig.~\ref{fig:crystal_shapes}(B1) and that in Fig.~\ref{fig:crystal_shapes}(B3), because the corner charge formulas for the other vertex-transitive polyhedra can be calculated from these results. 
It is similar to our previous work on cubic SGs \cite{PhysRevB.105.045126,PhysRevB.109.085114}, where it is enough to consider only three types of the crystal shapes (cube, octahedron, and tetrahedron) in order to obtain the corner charge formulas for all of the vertex-transitive polyhedra. 
The summary of the corner charge formulas in the tetragonal SGs are shown in Tab.~\ref{tab:corner_charge_tetra}. 

We note that in some SGs, the corner charge formulas are always trivial $(Q_{\text{corner}} = 0)$. 
Actually, in these SGs, both the hinge charge density and corner charge are proportional to the same charge imbalance $\Delta \omega$ at one certain WP, and therefore, the corner charge formulas are always zero under the hinge charge neutrality condition. 

\begin{figure*}
    \centering
    \includegraphics[scale = 0.6]{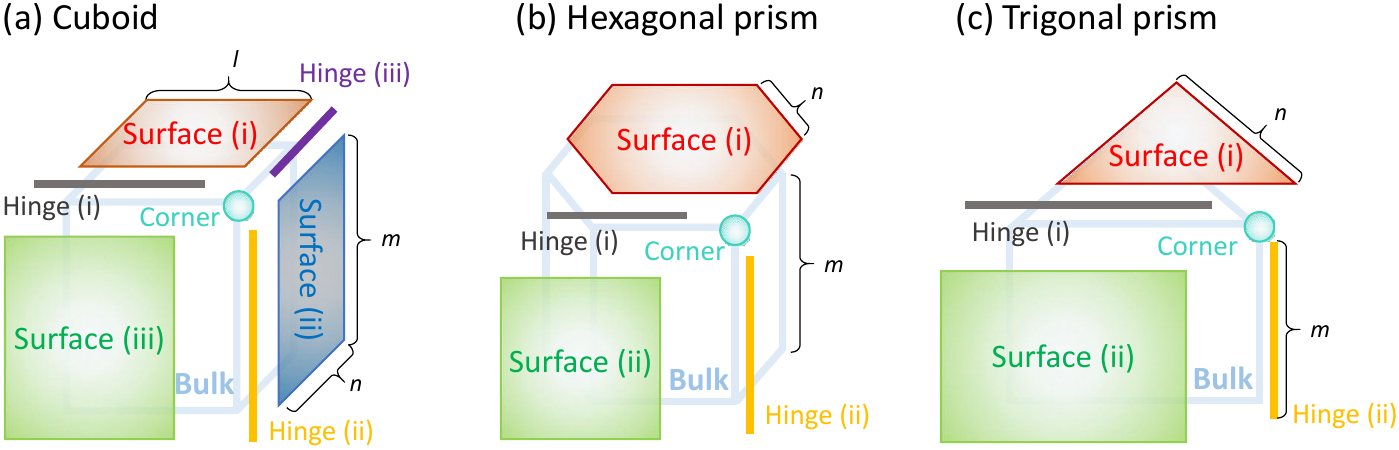}
    \caption{Divisions of the crystal into bulk, surface, hinge, and corner regions for crystal shapes of the (a) cuboid, (b) hexagonal prism, and (c) trigonal prism, for calculations of ambiguities in hinge and corner charges in orthorhombic, hexagonal, and trigonal SGs, respectively. The thicknesses of the surface, hinge, and corner regions are chosen to be sufficiently thick, so that the electronic states and nuclei positions in the bulk region are perfectly periodic.}
    \label{ambig}
\end{figure*}

\subsubsection{Hexagonal}
We also calculate the corner charge formulas for the hexagonal crystal shapes in the same way as the orthorhombic and tetragonal cases. 
We need to calculate the corner charges for hexagonal prisms in Figs.~\ref{fig:crystal_shapes}(C1) and (C3).
The list of the corner charge formulas are shown in Tab.~\ref{tab:corner_charge_hexa}. 
In Tab.~\ref{tab:corner_charge_hexa}, we find that in some SGs, the real-space formulas of the corner charge are always trivial for the same reason as in the tetragonal case.  

\subsection{Ambiguities in hinge charges and corner charges}
In Sec.~\ref{Sec.3}B, we derived the hinge and corner charges from the bulk electronic states and ionic positions by considering perfect crystals. 
In reality, the positions of the ions and the electronic Wannier centers might be modulated near the surfaces, hinges, and corners by relaxations near the crystal boundaries, which will affect the charge densities of the surfaces, hinges, and corners. 
Here, we show that possible relaxation of electronic states and ionic positions near the boundaries gives rise to ambiguities in the hinge charge density and the corner charge modulo some unit of charge. 
We explain that under certain conditions, these ambiguities can be calculated as filling anomalies for the surfaces and hinges by following Ref.~\cite{PhysRevB.105.045126}. 
In the following, we discuss the four cases, i.e. the orthorhombic, tetragonal, hexagonal, and trigonal cases separately.

First, let us consider the ambiguities for the crystal shape of a cuboid in the orthorhombic class as an example. 
The effects of surface relaxations can be studied by dividing the orthorhombic crystal shape into the bulk, surface, hinge, and corner regions as shown in Fig.~\ref{ambig}(a). 
Here, the surface region should be thick enough to confine the effect of relaxation within the surface region. 
Then we calculate the filling anomaly for each surface and hinge region up to the first order terms of $l$,$m$, and $n$: 
\begin{align}
    \text{Surface~(S-i)}:&\ lL_{1} + nN_{1} + C_{1},
    \label{eq:surface_part} \\
    \text{Surface~(S-ii)}:&\ mM_{1} + nN_{2} + C_{2}, \\
    \text{Surface~(S-iii)}:&\ lL_{2} + mM_{2} + C_{3}, \\
    \text{Hinge~(H-i)}:&\ lL_{3} + C_{4}, \\
    \text{Hinge~(H-ii)}:&\ mM_{3} + C_{5}, \\
    \text{Hinge~(H-iii)}:&\ nN_{3} + C_{6},
    \label{eq:hinge_part}
\end{align}
where $L_{i}, M_{i}, N_{i}\ (i = 1, 2, 3)$, and $C_{j}\ (j = 1 \cdots 6)$ are integers depending on the relaxation near the surfaces. 
To derive these results, we assumed that the crystallographic symmetries (including translation symmetries) of the surfaces and hinges fully follow those of the bulk. 
Now we specifically derive an ambiguity in the hinge charge density $\Delta \lambda^{l}_{\text{hinge}}$ for the Hinge (H-i) as an example. 
By focusing on the $l$-dependent terms in Eqs.~(\ref{eq:surface_part})--(\ref{eq:hinge_part}), a cuboid crystal contains two Surfaces (S-i), two Surfaces (S-iii) and four Hinges (H-i).
Therefore, the filling anomaly proportional to $l$ is given by $l(2L_{1} + 2L_{2} + 4L_{3})$. 
This filling anomaly induces a change in the hinge charge on the four Hinges (H-i), and therefore, the ambiguity in the hinge charge density $\Delta \lambda^{l}_{\text{hinge}}$ for the Hinge (H-i) is expressed as 
\begin{align}
    \Delta \lambda_{\text{hinge}}^{l} =& -\abs{e}\left( L_{3} + \frac{L_{1} + L_{2}}{2} \right). 
\end{align}
In the same way as $\Delta \lambda_{\text{hinge}}^{l}$, the ambiguities in the hinge charge densities $\Delta \lambda_{\text{hinge}}^{m}$ and $\Delta \lambda_{\text{hinge}}^{n}$ are given by 
\begin{align}
    \Delta \lambda_{\text{hinge}}^{m} =& -\abs{e}\left( M_{3} + \frac{M_{1} + M_{2}}{2} \right), \\
    \Delta \lambda_{\text{hinge}}^{n} =& -\abs{e}\left( N_{3} + \frac{N_{1} + N_{2}}{2} \right),
\end{align}
respectively. 
Therefore, the hinge charge densities $\lambda_{\text{hinge}}^{i}~(i = l,m,n)$ can be determined in terms of modulo $\abs{e}/2$. 
In the same way, the ambiguitiy in the corner charge $\Delta Q_{\text{corner}}$ is given by 
\begin{align}
   \Delta Q_{\text{corner}} = -\abs{e}\left( \frac{C_{1} + C_{2} + C_{3}}{4} + \frac{C_{4} + C_{5} + C_{6}}{2} \right).
\end{align}
Thus the corner charge $Q_{\text{corner}}$ can be determined in terms of modulo $\abs{e}/4$. 
These results support the validity of Eqs.~(\ref{eq:hinge_l_ortho})--(\ref{eq:corner_ortho}). 

Second, we study the square prism in the tetragonal class, which can be obtained by setting $l=n$ in the orthorhombic cases, and the hinge charge densities $\lambda_{\text{hinge}}^{l(=n)}$ and $\lambda_{\text{hinge}}^{m}$, and the corner charge $Q_{\text{corner}}$ can be generally determined in terms of modulo $\abs{e}/2$, $\abs{e}$, and $\abs{e}/4$, respectively. 
Here we note that there are some SGs, whose ambiguities in the hinge charge densities are different from the above values due to the configuration of the WPs in the surface and hinge regions. 
For example, in the SG No.~124 ($P$4/$mcc$), $\lambda_{\text{hinge}}^{m}$ is determined in terms of modulo $2\abs{e}$. 
However, even in such cases, an ambiguity in the corner charge is still defined in terms of modulo $\abs{e}/4$ because the modulation of the hinge charge density by an integer multiple of $|e|$ changes the corner charge only by integer multiple of $\abs{e}/2$.  
To summarize, we conclude that the corner charge formulas are robust against the boundary relaxation. 

Third, we consider the ambiguities in hinge and corner charges for the hexagonal prism in the hexagonal class. 
The conceptual picture of the region division is shown in Fig.~\ref{ambig}(b). 
In the same way as the orthorhombic and tetragonal cases, we obtain the ambiguities in the hinge charge densities and corner charge by considering the filling anomaly: 
\begin{align}
    \Delta \lambda_{\text{hinge}}^{m} =& -\abs{e}\left( M_{1} + M_{2} \right), \\
    \Delta \lambda_{\text{hinge}}^{n} =& -\abs{e}\left( N_{3} + \frac{N_{1} + N_{2}}{2} \right), \\
    \Delta Q_{\text{corner}} =& -\abs{e}\left( C_{3} + \frac{C_{2} + C_{4}}{2} + \frac{C_{1}}{6} \right), 
\end{align}
where $M_{i}~(i = 1, 2)$, $N_{j}~(j = 1, 2, 3)$, and $C_{k}~(k = 1, \cdots 4)$ are integers, and defined in the same way as the orthorhombic case. 
Thus the hinge charge densities $\lambda^{m}_{\text{hinge}}$ and $\lambda^{n}_{\text{hinge}}$, and corner charge $Q_{\text{corner}}$ can be determined in terms of modulo $\abs{e}$, $\abs{e}/2$, and $\abs{e}/6$, respectively. 
In this case, the corner charge in the hexagonal prism is defined in terms of modulo $\abs{e}/6$ from the bulk information, and therefore, the corner charge formulas are proved to be robust against the boundary relaxation. 

Finally, we explain why the corner charge in the trigonal crystal shapes is always trivial. 
Like the other crystal shapes, the crystal shape of the trigonal prism can be divided as shown in Fig.~\ref{ambig}(c). 
We can calculate the filling anomaly for each region by using integers $M_{i}~(i = 1, 2)$, $N_{j}~(j = 1, 2, 3)$, and $C_{k}~(k = 1, \cdots 4)$, which are defined through the same process as the orthorhombic case. 
As a result, we obtain the ambiguities: 
\begin{align}
    \Delta \lambda_{\text{hinge}}^{m} =& -\abs{e}\left( M_{1} + M_{2} \right), \\
    \Delta \lambda_{\text{hinge}}^{n} =& -\abs{e}\left( N_{3} + \frac{N_{2}}{2} + \frac{N_{1}}{6} \right), \\
    \Delta Q_{\text{corner}} =& -\abs{e}\left( C_{3} + \frac{C_{2} + C_{4}}{2} + \frac{C_{1}}{3} \right). 
    \label{eq:delta_corner_trig}
\end{align}
As for the hinge charge densities, we find that $\lambda^{m}_{\text{hinge}}$ and $\lambda^{n}_{\text{hinge}}$ can be defined in terms of modulo $\abs{e}$ and $\abs{e}/6$, respectively. 
On the other hand, the modulo part of the corner charge formulas turns out to be mod $\abs{e}/6$ from Eq.~(\ref{eq:delta_corner_trig}) while from the bulk information, that of corner charge formulas found to be $0$ or $\frac{\abs{e}}{6}$ mod $\abs{e}/3$. 
Therefore, even the nontrivial corner charge $\frac{\abs{e}}{6}$ in the perfect crystal can be trivialized by the boundary relaxation in the trigonal crystal shapes. 
Therefore, the corner charge formulas in the trigonal crystal shapes are always trivial.  

In summary, we obtained the real-space formulas of the quantized corner charge for all the crystal shapes with all the cylindrical SGs, and found that these formulas are universally given by Eq.~(\ref{eq:corner_charge_207}), i.e. $\Delta a$ (mod~2) divided by the number of corners $N$.
This universal result in Eq.~(\ref{eq:corner_charge_207}) for all the space groups suggests that there might be a general argument leading to Eq.~(\ref{eq:corner_charge_207}). 
In fact in two dimensions, a general argument leading to Eq.~(\ref{eq:corner_charge_207}) is given in \cite{PhysRevB.103.205123}. 
Meanwhile, such a general argument is not known so far, and the reason for this universal result is left as a future work. 
We also realize that there are some SGs in which their corner charges are always trivial.

We note that the corner charge is always determined by $\Delta a$ (mod~2).
It is universally given in terms of modulo 2 in the cylindrical family, which is natural from the following argument.
Suppose we consider two crystals with an identical cylindrical shape.
If we connect them, they form one long cylinder with neutral hinges. 
In this process, two corner charges are summed to become trivial, which implies that the corner charge in the cylindrical class always has a $\mathbb{Z}_{2}$ property.

\section{Corner charge formula in $k$ space\label{Sec.4}}

\begin{figure}[b]
    \centering
    \includegraphics[scale = 0.6]{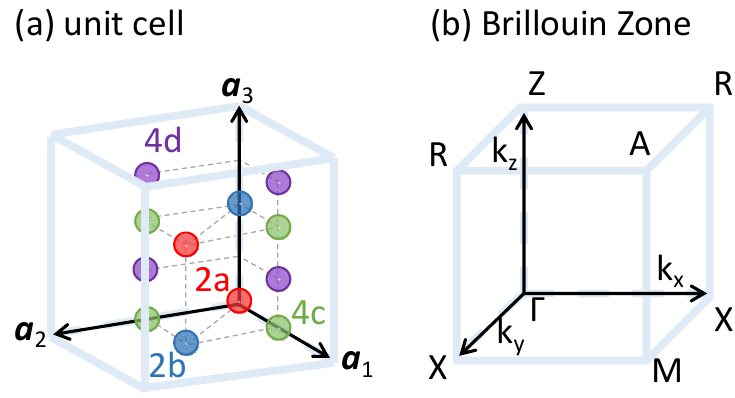}
    \caption{Calculation of the corner charge for the SG No.~128 ($P$4/$mnc$) in $k$ space. (a) Wyckoff positions (WPs) for the SG No.~128. The points with the same colors belong to the same WPs. The coordinates at the each WP are shown in Eq.~(\ref{eq:coordinate128}). The blue cuboid represents the unit cell. (b) HSPs in $k$ space for the SG No.~128. The blue cuboid represents 1/8 of the Brillouin zone. The $\Gamma$ point is the origin in $k$ space.}
    \label{SGNo.128}
\end{figure}

In this section, we construct corner charge formulas in spinless systems with the time-reversal symmetry (TRS) in terms of the electronic band structure. 
In spinful systems, because the electron numbers at each WP are always even due to the Kramers degeneracy \cite{PhysRevB.105.045126,PhysRevB.109.085114}, electrons do not contribute to corner charges, and the corner charges are determined only by the ionic charges i.e., $Q_{\text{corner}} = -\frac{\Delta a}{N}|e| = \frac{m_{a}}{N}|e|\ \ (\text{mod}\ \frac{2|e|}{N})$. 
Therefore, in the following, we focus on spinless cases. 

Before deriving the $k$-space formulas of the corner charge, we introduce the method of the elementary band representation (EBR) matrix according to Refs.~\cite{PhysRevB.105.045126,PhysRevB.109.085114,Po_2017, SI_SA_Watanabe,Bradlyn_2017,PhysRevB.103.165109}. 
The EBR matrix is an integer matrix that informs us of the relationship between the irreps of Wannier orbitals at the WPs in real space and irreps of filled bands at high-symmetry points (HSPs) in $k$ space. 
By using the EBR matrix method, we derive the number of Wannier orbitals at the WP $a$ from the number of filled bands at each HSP, which can be obtained from band structure calculations. 
Here we note that we are restricting our discussions to AIs, where the symmetry indicator of the set of filled bands is trivial \cite{Po_2017, SI_SA_Watanabe,Bradlyn_2017}. 
Moreover, we also assume that the Wannier center positions are uniquely determined from the set of Bloch wave functions of a given band structure. 

Here we explain the EBR matrix method, and show how we apply this method to derive the corner charge formulas.
Let $\bm{v}$ be a column vector consisting of the numbers of irreps for the filled bands at each HSP:
\begin{align}
    \bm{v} = (v_{\Pi}^{\rho_{1}},v_{\Pi}^{\rho_{2}},\cdots )^{t}, 
\end{align}
where $v_{\Pi}^{\rho}$ represents the number of filled bands with the irrep $\rho$ at HSP $\Pi$. 
Let $\bm{n}$ be a column vector consisting of the numbers of irreps of occupied Wannier orbitals at each maximal WP:
\begin{align}
    \bm{n} = (n_{\omega}^{\rho_{1}},n_{\omega}^{\rho_{2}},\cdots)^{t},
\end{align}
where $n_{\omega}^{\rho}$ is the number of the occupied Wannier orbital with the irrep $\rho$ at WP $\omega$. 
Then the EBR matrix $M$ is defined by
\begin{align}
    \bm{v} = M \bm{n}. 
    \label{eq:EBR}
\end{align}
One can calculate $M$ for each space group.

Our goal is to derive the value of $n^{\rho}_{1a}$ in the vector $\bm{n}$ from a given $\bm{v}$ by using Eq.~(\ref{eq:EBR}). 
It is not straightforward in general because $M$ is not invertible in general.
To achieve it, we consider the Smith decomposition of $M$: 
\begin{align}
    M = U^{-1}\Lambda V^{-1},
    \label{eq:Smith}
\end{align}
where $\Lambda$ is an integer diagonal matrix, whose matrix rank is $N$, in the form of $\Lambda_{ij}= \lambda\delta_{ij}$ ($\lambda_{i} > 0$ for $i = 1,\dots,N$ and $\lambda_{i} = 0$ for $i > N$), and $U$ and $V$ are  unimodular. 
From Eq.~(\ref{eq:Smith}), the vector $\bm{n}$ can generally be written as  
\begin{align}
    \bm{n} = V \Lambda^{p} U \bm{v} + V \bm{n_{0}},
    \label{eq:general_n}
\end{align}
where $\Lambda^{p}$ is the pseudoinverse matrix of $\Lambda$ (i.e., $\Lambda^{p}$ is the
diagonal matrix whose entry is $\lambda^{-1}_{i}$ for $i = 1,\dots,N$ and $0$ for $i > N$), and $\bm{n}_{0}$ is an arbitrary integer vector satisfying $\Lambda \bm{n}_{0} = 0$. 
Due to the ambiguity arising from the second term in Eq.~(\ref{eq:general_n}), $n_{i}$ can be determined in terms of modulo gcd($V_{ij}|_{j > N}$), where gcd indicates the greatest common divisor.
Therefore, the number of the Wannier orbitals at the Wyckoff position $\omega$ is given by 
\begin{equation}
    \begin{aligned}[b]
        n_{\omega} \equiv& \sum_{i \in \omega} \text{dim}(\rho_{i})(V \Lambda^{p} U \bm{v})_{i} \\
        &\left(\text{mod gcd}\left(\sum_{i \in \omega} \text{dim}(\rho_{i}) V_{i j}\right)_{j > N} \right),
        \label{eq:n_omega}
    \end{aligned}
\end{equation}
where $\rho_{i}$ represents the $i$th irrep in the vector $\bm{n}$, and $\sum_{i \in \omega}$ runs over all the irreps $i$ at the Wyckoff position $\omega$.  
In Sec.~\ref{Sec.3}, we learned that the filling anomaly in the cylindrical crystal shapes is universally given by $\Delta a~(=n_{a}-m_{a})$ in terms of modulo 2 to determine the corner charge. 
Therefore, we need to determine $n_{a}$ in Eq.~(\ref{eq:n_omega}) in terms of modulo 2. 

Here, we focus on the spinless system with the SG No.~128 ($P$4/$mnc$) as an example. 
We show the Wyckoff positions and HSPs in the SG No.~128 in Figs.~\ref{SGNo.128}(a) and (b), respectively. 
There are four types of the Wyckoff positions (i.e., $2a$, $2b$, $4c$, and $4d$) in the unit cell, and the coordinates at each WP are 
\begin{equation}
    \begin{aligned}[b]
        2a :&\ (0, 0, 0),\ \frac{1}{2}(\bm{a}_{1} + \bm{a}_{2} + \bm{a}_{3}), \\
        2b :&\ \frac{1}{2}(\bm{a}_{1} + \bm{a}_{2}),\ \frac{1}{2}\bm{a}_{3}, \\
        4c :&\ \frac{1}{2}\bm{a}_{1},\ \frac{1}{2}(\bm{a}_{1} + \bm{a}_{3}), \\
        &\ \frac{1}{2}\bm{a}_{2},\ \frac{1}{2}(\bm{a}_{2} + \bm{a}_{3}), \\
        4d :&\ \frac{1}{4}(2\bm{a}_{1} + \bm{a}_{3}),\ \frac{1}{4}(2\bm{a}_{1} + 3\bm{a}_{3}), \\
        &\ \frac{1}{4}(2\bm{a}_{2} + \bm{a}_{3}),\ \frac{1}{4}(2\bm{a}_{2} + 3\bm{a}_{3}),
        \label{eq:coordinate128}
    \end{aligned}
\end{equation}
where $\bm{a}_{i}~(i=1,2,3)$ are the primitive lattice vectors in the orthorhombic unit cell.
The site-symmetry groups of $2a$, $2b$, $4c$, and $4d$ are isomorphic to the point groups $C_{4h}$, $C_{4h}$, $C_{2h}$, and $D_{2}$, respectively. 
On the other hand, there are six kinds of HSPs called $A$, $\Gamma$, $M$, $Z$, $R$, and $X$ in the SG No.~128. 
The little groups at $A$, $\Gamma$, $M$, and $Z$ are isomorphic to the point group $D_{4h}$, and those at $R$ and $X$ are isomorphic to the point group $D_{2h}$.

Now we consider the EBR matrix for the SG No.~128. 
Band representations for the Wannier orbitals located at all the maximal Wyckoff positions are expressed as a vector $\bm{v}$ in the form
\begin{equation}
    \begin{aligned}[b]
        (v^{1}_{A}, v^{2}_{A}, v^{3}_{A}, v^{4}_{A}, v^{1+}_{\Gamma}, v^{1-}_{\Gamma}, v^{2+}_{\Gamma}, v^{2-}_{\Gamma}, v^{3+}_{\Gamma}, v^{3-}_{\Gamma}, v^{4+}_{\Gamma}, \\ v^{4-}_{\Gamma}, v^{5+}_{\Gamma}, v^{5-}_{\Gamma}, v^{1+}_{M}, v^{1-}_{M}, v^{2+}_{M}, v^{2-}_{M}, v^{3+}_{M}, v^{3-}_{M}, v^{4+}_{M}, \\ v^{4-}_{M}, v^{5+}_{M}, v^{5-}_{M}, v^{1}_{Z}, v^{2}_{Z}, v^{3}_{Z}, v^{4}_{Z}, v^{1+}_{R}, v^{1-}_{R}, v^{1}_{X}, v^{2}_{X} 
        )^{t}.
        \label{eq:basis_v}
    \end{aligned}
\end{equation} 
In this case, the notation of the each irrep in Eq.~(\ref{eq:basis_v}) follows the Bilbao Crystallographic Server \cite{Aroyo:xo5013}.
We are restricting ourselves to the AIs, which are band insulators with a trivial symmetry indicator, meaning that its vector $\bm{v}$ is written as a linear combination of the column vectors of the EBR matrix with integer coefficients. 
The coefficients form a vector $\bm{n}$ in the form 
\begin{equation}
    \begin{aligned}[b]
        (n^{A_{g}}_{2a}, n^{A_{u}}_{2a}, n^{B_{g}}_{2a}, n^{B_{u}}_{2a}, n^{E_{g}}_{2a}, n^{E_{u}}_{2a}, n^{A_{g}}_{2b}, n^{A_{u}}_{2b}, n^{B_{g}}_{2b}, n^{B_{u}}_{2b}, \\
        n^{E_{g}}_{2b}, n^{E_{u}}_{2b}, n^{A_{g}}_{4c}, n^{A_{u}}_{4c}, n^{B_{g}}_{4c}, n^{B_{u}}_{4c}, n^{A}_{4d}, n^{B_{1}}_{4d}, n^{B_{2}}_{4d}, n^{B_{3}}_{4d})^{t}.
    \end{aligned}
\end{equation}
Here $E_{g}$ and $E_{u}$ represent $E^{1}_{g} \oplus E^{2}_{g}$ and $E^{1}_{u} \oplus E^{2}_{u}$, respectively due to TRS. 
In this case, we can construct the EBR matrix based on topological quantum chemistry \cite{Bradlyn_2017} and theory of symmetry-based indicators \cite{Po_2017,SI_SA_Watanabe} by using the Bilbao Crystallographic Server \cite{Aroyo:xo5013} as follows: 
\begin{widetext}
    \begin{align}
        M = \mqty(1&0&0&1&0&0&0&1&1&0&0&0&0&0&1&1&0&0&1&1 \\
                  0&1&1&0&0&0&1&0&0&1&0&0&0&0&1&1&0&0&1&1 \\
                  0&0&0&0&1&1&0&0&0&0&1&1&1&1&0&0&1&1&0&0 \\
                  0&0&0&0&1&1&0&0&0&0&1&1&1&1&0&0&1&1&0&0 \\
                  1&0&0&0&0&0&1&0&0&0&0&0&1&0&0&0&1&0&0&0 \\
                  0&1&0&0&0&0&0&1&0&0&0&0&0&1&0&0&1&0&0&0 \\
                  0&0&1&0&0&0&0&0&1&0&0&0&1&0&0&0&0&1&0&0 \\
                  0&0&0&1&0&0&0&0&0&1&0&0&0&1&0&0&0&1&0&0 \\
                  1&0&0&0&0&0&1&0&0&0&0&0&1&0&0&0&0&1&0&0 \\
                  0&1&0&0&0&0&0&1&0&0&0&0&0&1&0&0&0&1&0&0 \\
                  0&0&1&0&0&0&0&0&1&0&0&0&1&0&0&0&1&0&0&0 \\
                  0&0&0&1&0&0&0&0&0&1&0&0&0&1&0&0&1&0&0&0 \\
                  0&0&0&0&2&0&0&0&0&0&2&0&0&0&2&0&0&0&1&1 \\
                  0&0&0&0&0&2&0&0&0&0&0&2&0&0&0&2&0&0&1&1 \\
                  0&0&0&0&1&0&0&0&0&0&1&0&0&1&0&0&0&1&0&0 \\
                  0&0&0&0&0&1&0&0&0&0&0&1&1&0&0&0&0&1&0&0 \\
                  0&0&0&0&1&0&0&0&0&0&1&0&0&1&0&0&1&0&0&0 \\
                  0&0&0&0&0&1&0&0&0&0&0&1&1&0&0&0&1&0&0&0 \\
                  0&0&0&0&1&0&0&0&0&0&1&0&0&1&0&0&1&0&0&0 \\
                  0&0&0&0&0&1&0&0&0&0&0&1&1&0&0&0&1&0&0&0 \\
                  0&0&0&0&1&0&0&0&0&0&1&0&0&1&0&0&0&1&0&0 \\
                  0&0&0&0&0&1&0&0&0&0&0&1&1&0&0&0&0&1&0&0 \\
                  1&0&1&0&0&0&1&0&1&0&0&0&0&0&0&2&0&0&1&1 \\
                  0&1&0&1&0&0&0&1&0&1&0&0&0&0&2&0&0&0&1&1 \\
                  0&0&1&1&0&0&0&0&1&1&0&0&1&1&0&0&1&1&0&0 \\
                  1&1&0&0&0&0&1&1&0&0&0&0&1&1&0&0&1&1&0&0 \\
                  0&0&0&0&1&1&0&0&0&0&1&1&0&0&1&1&0&0&1&1 \\
                  0&0&0&0&1&1&0&0&0&0&1&1&0&0&1&1&0&0&1&1 \\
                  1&0&1&0&2&0&0&1&0&1&0&2&1&1&1&1&1&1&1&1 \\
                  0&1&0&1&0&2&1&0&1&0&2&0&1&1&1&1&1&1&1&1 \\
                  0&1&0&1&2&0&0&1&0&1&2&0&0&2&2&0&1&1&1&1 \\
                  1&0&1&0&0&2&1&0&1&0&0&2&2&0&0&2&1&1&1&1 ).
    \end{align}
Here the $(i, j)$ component of $M$ indicates the number of times the $i$th irrep in the $k$ space appears in the EBR induced from the $j$th irrep in the real space. 
We can derive the matrices $V$ and $\Lambda$ in Eq.~(\ref{eq:Smith}) by carrying out the Smith decomposition:
\begin{align}
    V = \mqty(1&0&0&1&0&1&2&1&-2&2&0&-1&1&-1&1&-1&0&-1&-1&0 
               \\
              0&1&0&0&1&0&1&0&0&0&0&0&0&-1&0&0&0&-1&-1&0 \\
              0&0&0&0&-1&1&0&0&0&1&0&-1&0&-1&1&-1&0&-1&0&0 \\
            0&0&0&-1&0&-1&-2&-2&4&-2&0&0&-1&-1&0&0&0&-1&0&0 \\
              0&0&1&0&0&0&1&0&1&0&1&1&-1&0&-1&1&1&0&0&0 \\
              0&0&0&0&0&0&0&0&0&0&0&0&0&0&0&0&1&0&0&0 \\
              0&0&0&0&0&-1&-1&-1&2&-1&0&1&-1&0&0&0&0&0&1&0 \\
              0&0&0&0&0&0&0&0&0&0&0&0&0&0&0&0&0&0&1&0 \\
              0&0&0&0&0&0&0&0&0&0&0&1&0&0&0&0&0&0&0&0 \\
              0&0&0&0&0&0&0&0&0&0&0&0&1&0&0&0&0&0&0&0 \\
              0&0&0&0&0&0&0&0&0&0&-1&-1&1&-1&1&-1&-1&-1&0&0 \\
              0&0&0&0&0&0&0&0&-1&1&0&0&0&-1&1&-1&-1&-1&0&0 \\
              0&0&0&0&0&0&0&0&0&-1&0&0&0&1&-1&1&0&0&0&0 \\
              0&0&0&0&0&0&0&0&0&0&0&0&0&1&0&0&0&0&0&0 \\
              0&0&0&0&0&0&0&0&0&0&0&0&0&0&1&0&0&0&0&0 \\
              0&0&0&0&0&0&0&0&0&0&0&0&0&0&0&1&0&0&0&0 \\
              0&0&0&0&0&0&-1&0&0&0&0&0&0&0&0&0&0&1&0&0 \\
              0&0&0&0&0&0&0&0&0&0&0&0&0&0&0&0&0&1&0&0 \\
              0&0&0&0&0&0&0&1&-2&0&0&0&0&2&-2&0&0&2&0&-1 \\ 
              0&0&0&0&0&0&0&0&0&0&0&0&0&0&0&0&0&0&0&1 ),
\end{align}
\begin{align}
    \lambda_{1} = \lambda_{2} = \cdots = \lambda_{10} = 1, \lambda_{11} = 2, N = 11. 
\end{align}
From Eq.~(\ref{eq:n_omega}), the ambiguity of $n_{2a}$ from the symmetry indicator can be calculated as 
\begin{equation}
    \begin{aligned}[b]
        \text{gcd}\left(\sum_{i = 1, 2, 3, 4} \text{dim}(\rho_{i}) V_{i j}\right)_{j > 11} =& \text{gcd}\left[(1,1,1,1)\mqty(-1&1&-1&1&-1&0&-1&-1&0 
               \\0&0&-1&0&0&0&-1&-1&0 \\-1&0&-1&1&-1&0&-1&0&0 \\0&-1&-1&0&0&0&-1&0&0 \\)\right] \\
        =& \text{gcd}(-2, 0, -4, 2, -2, 0, -4, -2, 0) = 2,
        \label{eq:SG128_ambi}
    \end{aligned}
\end{equation}
where the irreps $\rho_{1}$, $\rho_{2}$, $\rho_{3}$, and $\rho_{4}$ represent one-dimensional irreps $A_{g}$, $A_{u}$, $B_{g}$, and $B_{u}$, respectively. 
The dimensions of $\rho_{5}~(= E_{g})$ and $\rho_{6}~(= E_{u})$ are even numbers, and thus we exclude their contribution to the ambiguity in the above equation because they do not contribute to the corner charge in Eq.~(\ref{eq:corner_charge_207}).  
\end{widetext} 

\begin{figure}[b]
    \centering
    \includegraphics[scale = 0.6]{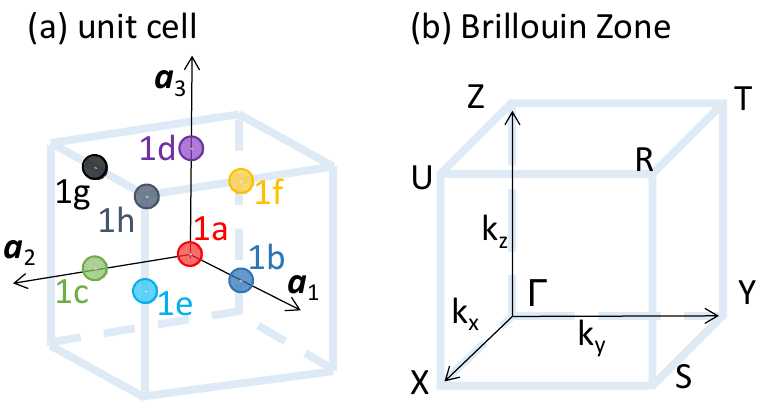}
    \caption{(a) Wyckoff positions (WPs) in the SG No.~16 ($P$222). $\bm{a}_{i}\ (i = 1, 2, 3)$ are primitive lattice vectors in the cubic unit cell. The blue cube represents the unit cell. (b) HSPs in the $k$ space for the SG No.~16. The blue cube represents 1/8 of the Brillouin zone. The $\Gamma$ point is the origin in $k$ space.}
    \label{SGNo.16}
\end{figure}

In Sec.~\ref{Sec.3}, we found that to determine the corner charge (Eq.~(\ref{eq:corner_charge_207})), we need to calculate the filling anomaly $\Delta a~(= n_{a}-m_{a})$ in terms of modulo 2, which is the same as the ambiguity of $n_{2a}$ in Eq.~(\ref{eq:SG128_ambi}) in the $k$-space formula. 
Therefore, we can safely establish the $k$-space formula of the corner charge for the SG No.~128. 
From Eq.~(\ref{eq:general_n}), $n_{2a}$ is given by 
\begin{align}
    n_{2a} \equiv v^{1}_{A} + v^{3}_{A} + v^{1-}_{\Gamma} + v^{2+}_{\Gamma} + v^{3}_{Z} - m_{2a}\ \ (\text{mod}\ 2). 
\end{align}
Similarly we can calculate $k$-space formulas of the filling anomaly $\Delta a (= n_{a} - m_{a})$ for all the cylindrical SGs as summarized in Tab.~\ref{Tab:k-space_filling_anomaly}. 

\begin{table}
\caption{Summary for the $k$-space formulas of the filling anomaly with TRS in the cylindrical crystal shapes. The corner charge formulas are given by dividing their filling anomaly by $N$ and multiplying by $-\abs{e}$. }
\centering
\renewcommand{\arraystretch}{1.2}
\begin{ruledtabular}
\begin{tabular}{cc}
    SG number & $\Delta a_{\text{spinless}}$ (mod 2) \\\hline
    47 & $v_{\Gamma}^{2-} + v_{\Gamma}^{3-} + v_{\Gamma}^{4-} + v_{R}^{2+} + v_{R}^{3+} + v_{R}^{4+}$ \\
    & $+ v_{S}^{1-} + v_{T}^{1-} + v_{U}^{1-} + v_{X}^{1+} + v_{Y}^{1+} + v_{Z}^{1+} - m_{a}$ \\
    65 & $v_{\Gamma}^{1+} + v_{\Gamma}^{4+} + v_{\Gamma}^{4-} + v_{T}^{1+} + v_{T}^{3+} + v_{T}^{3-}$ \\
    & $ + v_{Y}^{1+} + v_{Z}^{1+} + v_{R}^{1-} + v_{S}^{1+} - m_{a}$ \\
    69 & $v_{\Gamma}^{1+} + v_{\Gamma}^{2+} + v_{\Gamma}^{2-} + v_{\Gamma}^{3-} + v_{\Gamma}^{4-} + v_{T}^{2+} $ \\ 
    & $ + v_{Y}^{1-} + v_{Z}^{1+} + v_{L}^{1+} - m_{a}$ \\
    71 & $v_{\Gamma}^{2-} + v_{\Gamma}^{3-} + v_{\Gamma}^{4-} + v_{X}^{1+} + v_{R}^{1+} + v_{S}^{1+} $ \\
    & $ + v_{T}^{1+} + v_{W}^{1} - m_{a}$ \\
    83 & $v_{A}^{2-} + v_{A}^{3-} + v_{\Gamma}^{2+} + v_{\Gamma}^{3-} + v_{M}^{1+} + v_{M}^{3+} $ \\
    & $ + v_{Z}^{1-} + v_{Z}^{3+} - m_{a}$ \\
    87 & $ v_{\Gamma}^{1+} + v_{\Gamma}^{1-} + v_{\Gamma}^{2+} + v_{\Gamma}^{3-} + v_{M}^{1+} + v_{M}^{3+}$\\
    & $ + v_{P}^{1} + v_{P}^{3} - m_{a} $ \\
    89 & $v_{A}^{3} + v_{A}^{4} + v_{\Gamma}^{1} + v_{\Gamma}^{2} + \xi - m_{a} $\\
    97 & $ \xi - m_{a} $ \\
    111 & $v_{A}^{2} + v_{A}^{3} + v_{A}^{4} + v_{\Gamma}^{2} + v_{M}^{2} + v_{Z}^{1} $ \\
    & $ + v_{R}^{1} + v_{X}^{1} - m_{a} $ \\
    121 & $v_{\Gamma}^{1} + v_{\Gamma}^{3} + v_{\Gamma}^{4} + v_{M}^{1} + v_{P}^{1} + v_{P}^{2} - m_{a}$  \\
    123 & $v_{A}^{3+} + v_{A}^{4+} + v_{\Gamma}^{1-} + v_{\Gamma}^{2-} + v_{M}^{3-} + v_{M}^{4-} $ \\ 
    & $+ v_{Z}^{1+} + v_{Z}^{2+} - m_{a}$ \\
    124 & $v_{A}^{1} + v_{A}^{2} + v_{\Gamma}^{1-} + v_{\Gamma}^{2-} + v_{M}^{1+} + v_{M}^{2+} - m_{a}$ \\
    125 & $v_{A}^{1} + v_{A}^{2} + v_{\Gamma}^{1+} + v_{\Gamma}^{2-} + v_{Z}^{1-} + v_{Z}^{2+} - m_{a}$ \\
    126 & $v_{A}^{2} + v_{\Gamma}^{1-} + v_{\Gamma}^{2+} - m_{a}$ \\
    127 & $v_{A}^{1+} + v_{A}^{1-} + v_{A}^{2+} + v_{A}^{2-} + v_{A}^{5+} + v_{\Gamma}^{1+} $ \\
    & $ + v_{\Gamma}^{3+} + v_{\Gamma}^{5+} + v_{Z}^{1+} + v_{Z}^{2+} - m_{a}$ \\
    128 & $v_{A}^{1} + v_{A}^{3} + v_{\Gamma}^{1-} + v_{\Gamma}^{2+} + v_{Z}^{3} - m_{a}$ \\
    131 & $v_{A}^{1} + v_{A}^{2} + v_{A}^{4} + v_{\Gamma}^{1+} + v_{\Gamma}^{2-} + v_{\Gamma}^{3+} $ \\
    & $ + v_{\Gamma}^{3-} + v_{M}^{1+} + v_{M}^{2+} + v_{Z}^{3} + v_{R}^{1+} + v_{R}^{1-} - m_{a}$ \\
    132 & $v_{A}^{1} + v_{A}^{3} + v_{A}^{4} + v_{\Gamma}^{2-} + v_{\Gamma}^{3+} + v_{M}^{1+} $ \\
    & $ + v_{M}^{2+} + v_{R}^{1} - m_{a}$ \\
    134 & $v_{A}^{1+} + v_{A}^{2+} + v_{A}^{2-} + v_{A}^{3-} + v_{A}^{5+} + v_{\Gamma}^{1+} $ \\
    & $ + n_{\Gamma}^{3+} + n_{\Gamma}^{5+} - m_{a}$ \\
    136 & $v_{A}^{1} + v_{A}^{3} + v_{A}^{4} + v_{\Gamma}^{1+} + v_{\Gamma}^{1-} + v_{\Gamma}^{5+} + v_{R}^{1+} - m_{a}$ \\
    139 & $v_{\Gamma}^{3-} + v_{\Gamma}^{4+} + v_{M}^{3+} + v_{M}^{4-} + v_{P}^{1} + v_{P}^{2} - m_{a}$ \\
    140 & $v_{\Gamma}^{1-} + v_{\Gamma}^{2+} + v_{\Gamma}^{5+} + v_{\Gamma}^{5-} + v_{M}^{1-} + v_{M}^{2+} - m_{a}$ \\
    175 & $v_{A}^{1-} + v_{A}^{2-} + v_{\Gamma}^{1+} + v_{\Gamma}^{2+} - m_{a}$\\
    177 & $v_{A}^{1} + v_{A}^{2} + v_{A}^{4} + v_{A}^{5} + v_{\Gamma}^{3} + v_{K}^{2}$ \\
    & $+ v_{L}^{1} + \xi - m_{a}$ \\
    191 & $v_{A}^{2-} + v_{A}^{4+} + v_{A}^{4-} + v_{A}^{5-} + v_{\Gamma}^{2+} + v_{\Gamma}^{3+} $ \\
    & $+ v_{\Gamma}^{3-} + v_{\Gamma}^{5+} + v_{L}^{1-} + v_{M}^{1+} - m_{a}$ \\
    192 & $v_{A}^{5} + v_{A}^{6} + v_{\Gamma}^{1-} + v_{\Gamma}^{3+} + v_{\Gamma}^{3-} + v_{\Gamma}^{5+} + v_{M}^{1+} - m_{a}$ \\
\end{tabular}
\end{ruledtabular}
\label{Tab:k-space_filling_anomaly}
\end{table}

We note that for the SGs No.~89 ($P$422), 97 ($I$422), and 177 ($P$622), we cannot determine the $k$-space formulas of the filling anomaly in terms of modulo 2 only from the EBR matrix. 
In these SGs, the ambiguity in $n_{a}$ obtained from Eq.~(\ref{eq:general_n}) is modulo $1$, not modulo $2$.
Therefore from the EBR method, we cannot determine $n_{a}$ in terms of mod $2$. 
It means that the corner charge cannot be determined only from the irreps of the occupied bands at the HSPs. 
In these cases, when the TRS is preserved, we can calculate the corner charge in those cases by introducing the $\mathbb{Z}_{2}$ topological invariant $\xi~(=0, 1)$ for the SG No.~16 ($P$222) with TRS introduced in Refs.~\cite{PhysRevB.109.085114,Ono=Shiozaki_top_inv}, because the SG No.~16 is a subgroup of those SGs.  
We show the formula of the topological invariant $\xi$ in Appendix \ref{Ap:inv}. 
 
Now, we explain how to derive the $k$-space formulas of the filling anomaly for the above three SGs. 
First, we calculate the values of the $\mathbb{Z}_{2}$ invariant $\xi$ for each atomic insulator (AI) in the SG considered. 
As an example, here we show the values of $\xi$ for the SG No.~16. 
The Wyckoff positions and HSPs in the SG No.~16 are shown in Fig.~\ref{SGNo.16}. 
The site-symmetry groups of all the maximal Wyckoff positions are isomorphic to the point group $D_{2}$, whose irreps are listed in Tab.~\ref{tab:D2}. 
Then the values of the invariant $\xi$ for the SG No.~16 are  
\begin{align}
    \xi (n^{A_{1}}_{1h}) &= \xi (n^{B_{1}}_{1h}) = \xi (n^{B_{2}}_{1h}) = \xi (n^{B_{3}}_{1h}) = 1,
    \label{eq:non_trivial_xi}
\end{align}
\begin{align}
    \xi(n^{\rho}_{\omega}) &= 0 \ \text{otherwise}.
\end{align}
These results apply to the above three SGs, which are supergroups of the SG No.~16. 
Next, we define a pseudo-EBR matrix $\Tilde{M}$ from the EBR matrix $M$, by adding one row representing the values of $\xi$ modulo 2. 
In this case, we also incorporate $\xi$ into the form of $\bm{v}$, and define the vector $(\bm{v},\xi)$ as $\Tilde{\bm{v}}$. 
Then we apply the Smith decomposition to this pseudo-EBR matrix, and calculate $n_{a}$ just as before. 
As a result, we can obtain the $k$-space formulas of the filling anomaly for the SGs No.~89, 97, and 177 in terms of modulo 2 as shown in Tab.~\ref{Tab:k-space_filling_anomaly}. 

To summarize, we have derived the $k$-space formulas for filling anomaly and quantized corner charge for all the cases of cylindrical crystal shapes with TRS.

\begin{table}[H]
    \caption{Character table of the point group $D_{2}$.}
    \centering
    \begin{ruledtabular}
    \begin{tabular}{ccccc}
         $D_{2}$ & $E$ & $C_{2x}$ & $C_{2y}$ & $C_{2z}$ \\ \hline
         $A_{1}$ & 1 & 1 & 1 & 1 \\
         $B_{1}$ & 1 & $-1$ & $-1$ & 1 \\
         $B_{2}$ & 1 & $-1$ & 1 & $-1$ \\
         $B_{3}$ & 1 & 1 & $-1$ & $-1$ 
    \end{tabular}
    \end{ruledtabular}
    \label{tab:D2}
\end{table}

\section{Material examples}

\begin{figure*}
    \centering
    \includegraphics[scale = 0.5]{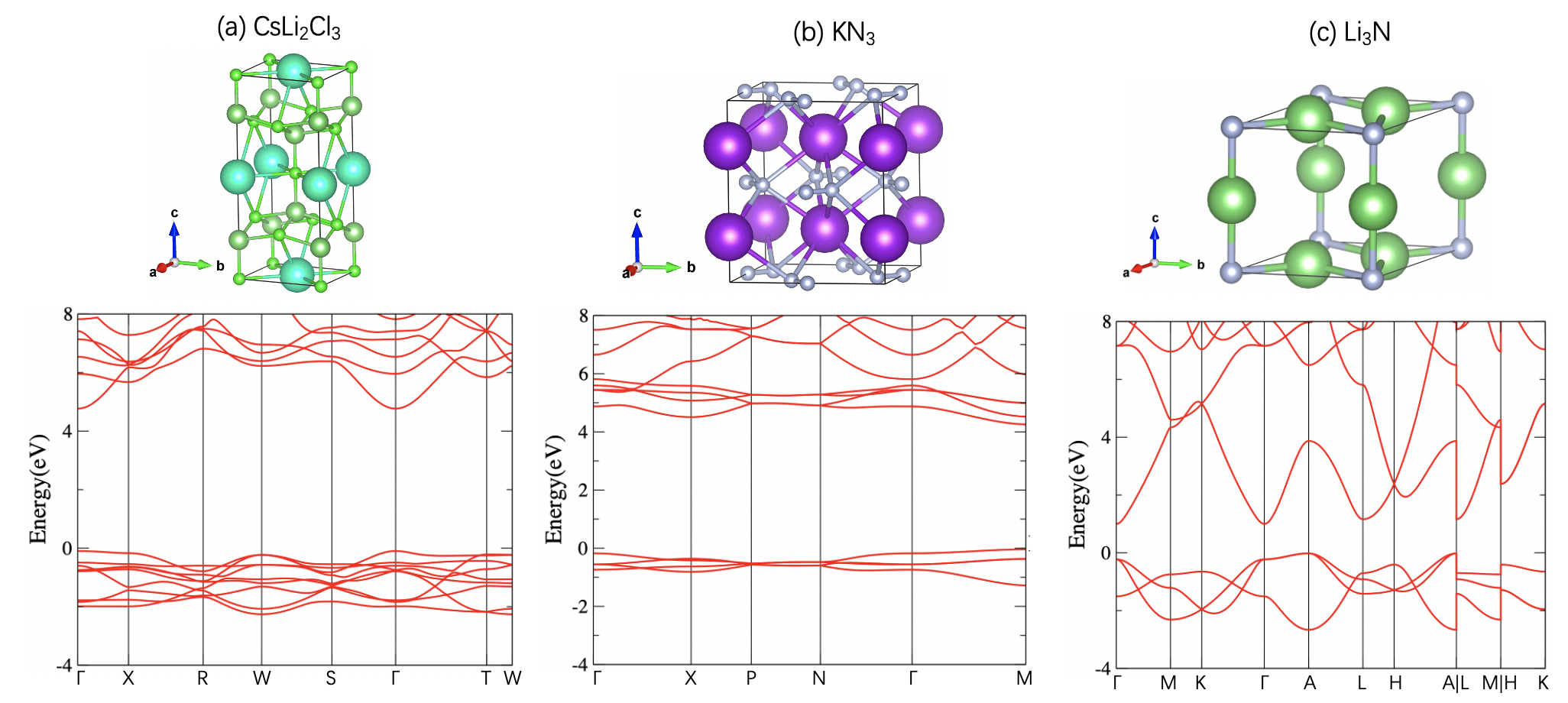}
    \caption{Crystal structures and spinless band structures for (a) CsLi$_2$Cl$_3$, (b) KN$_3$ and (c) Li$_3$N. All the Fermi energies are set to be 0. (a) CsLi$_2$Cl$_3$ belong to the SG No. 71 ($Immm$), with 36 valence electrons. Thus, there are 18 occupied bands below the Fermi energy in the spinless case. (b) KN$_3$ belong to the SG No. 140 ($I4/mcm$), with 24 valence electrons. Thus, there are 12 occupied bands below the Fermi energy in the spinless electronic band structure. (c) Li$_3$N belong to the SG No. 191 ($P6/mmm$), with 14 valence electrons. Thus, there are 7 occupied bands below the Fermi energy. }
    \label{fig:DFT}
\end{figure*}

Here we show the three candidate materials with a quantized corner charge (that is, $\text{CsLi}_{2}\text{Cl}_3$, $\text{KN}_{3}$, and $\text{Li}_{3}\text{N}$) by using $ab$ $initio$ calculations. 
The crystal structures and spinless band structures for the three materials are shown in Fig.~\ref{fig:DFT}. All the Fermi energies are set to be zero.
All these materials have the characteristic of ionic crystals. 
Calculations for the spinless electronic band structures in Fig.~\ref{fig:DFT} are  conducted with VASP, employing the generalized gradient approximation (GGA) of the Perdew–Burke–Ernzerhof (PBE)-type exchange-correlation functional~\cite{kresse1996efficient,kresse1994ab,kresse1996efficiency,kresse1993ab}. 
The pseudopotentials include 36, 24 and 14 valence electrons for $\text{CsLi}_{2}\text{Cl}_3$, $\text{KN}_{3}$, and $\text{Li}_{3}\text{N}$, respectively.
The cut-off energy for the plane wave basis set was determined by adding 25\% to the ENMAX value specified in the pseudopotential file. 
Furthermore, a Monkhorst-Pack $k$ point mesh centered on $\Gamma$ with a density of 30 points per $\AA^{-1}$ was utilized for self-consistent field calculations. Through the analysis of irreps, we have verified that each of the three materials possesses nontrivial corner charges as follows. 
In the calculation with VASP, the band structure for the valence electrons is calculated. 
Meanwhile, the core electrons are tightly bound to the nuclei, and the nuclei with the core electrons are considered as ions located at some WPs. 

\subsection{CsLi$_2$Cl$_3$} 
The crystal structure and spinless electronic band structure of $\text{CsLi}_{2}\text{Cl}_3$ crystals are illustrated in Fig.~\ref{fig:DFT}(a). 
In this crystal, Cl nuclei with 10 core electrons are located at WP $2a$ and WP $4f$, Cs nuclei with 36 core electrons are located at the WP $2b$, and Li nuclei are located in WP $4e$. 
Yet in our DFT calculations, we consider only 36 valence electrons in the primitive cell by choosing the pseudopotentials, resulting in 18 occupied bands in the spinless case. 
Additionally, this crystal belongs to space group No. 71 ($Immm$), and we examine a crystal shape resembling a cuboid in the orthorhombic class (refer to Subsection~\ref{Sec.3}A).
Thus, from our result in Tab.~\ref{Tab:k-space_filling_anomaly}, the corner charge formula $Q^{\text{No.~71}}_{\text{corner}}$ for the SG No.~71 is given by 
\begin{equation}
    \begin{aligned}[b]
        Q^{\text{No.~71}}_{\text{corner}} =& - \frac{\abs{e}}{8}(v_{\Gamma}^{2-} + v_{\Gamma}^{3-} + v_{\Gamma}^{4-} + v_{X}^{1+}\\
        & + v_{R}^{1+} + v_{S}^{1+} + v_{T}^{1+} + v_{W}^{1} - m_{a})
    \end{aligned}
\end{equation}
in terms of modulo $\abs{e}/4$. 
In this case, from the result obtained from the $ab$ $initio$ calculation with 18 spinless occupied bands, we find that $v_{\Gamma}^{2-}=3$, $v_{\Gamma}^{3-}=3$, $v_{\Gamma}^{4-}=5$, $v_{X}^{1+}=5$, $v_{R}^{1+}=5$, $v_{S}^{1+}=7$, $v_{T}^{1+}=7$, and $v_{W}^{1}=5$. 
Moreover, since there is a Cl nucleus with 10 core electrons located at the WP $a$, the total ionic charge $m_{a}$ located at WP $2a$ (per elementary charge) equals to $m_{a}=7$. 
Thus the fractional corner charge $Q^{\text{No.~71}}_{\text{corner}}$ for the $\text{CsLi}_{2}\text{Cl}_3$ crystal is 
\begin{align}
    Q^{\text{No.~71}}_{\text{corner}} = -\frac{\abs{e}}{8}\ \ \left(\text{mod}\ \frac{\abs{e}}{4}\right). 
\end{align}

\subsection{KN$_{3}$}
The second example is $\text{KN}_{3}$ crystals, whose crytsal structure and electronic band structure is shown in Fig.~\ref{fig:DFT}(b) belonging to the SG No~140 ($I$4/$mcm$). 
In our calculation on this crystal, K nuclei, each with 10 core electrons, are located at the WP $4a$ and N nuclei, each with 2 core electrons, are located at the WP $4d$ and WP $8h$. 
In the DFT calculations, only 24 valence electrons are considered in the primitive cell, thereby 12 occupied bands in the spinless electronic band structure. 
Here we assume that the crystal shape is a square prism in the tetragonal class, and its center is located at the WP $4a$.  
In this case, from Tab.~\ref{Tab:k-space_filling_anomaly}, the corner charge formula $Q^{\text{No.~140}}_{\text{corner}}$ for the SG No.~140 is given by 
\begin{equation}
    \begin{aligned}[b]
        Q^{\text{No.~140}}_{\text{corner}} =& -\frac{\abs{e}}{8}(v_{\Gamma}^{1-} + v_{\Gamma}^{2+} + v_{\Gamma}^{5+} \\
        & + v_{\Gamma}^{5-} + v_{M}^{1-} + v_{M}^{2+} - m_{a})
    \end{aligned}
\end{equation}
in terms of modulo $\abs{e}/4$. 
In the same way as the case of $\text{CsLi}_{2}\text{Cl}_3$, from the 24 spinless occupied bands, we find that $v_{\Gamma}^{1-}=1$, $v_{\Gamma}^{2+}=1$, $v_{\Gamma}^{5+}=2$, $v_{\Gamma}^{5-}=4$, $v_{M}^{1-}=2$, $v_{M}^{2+}=2$, and $m_{a}=9$. 
Thus the fractional corner charge $Q^{\text{No.~140}}_{\text{corner}}$ for the KN$_{3}$ crystal is 
\begin{align}
    Q^{\text{No.~140}}_{\text{corner}} = -\frac{\abs{e}}{8}\ \ \left(\text{mod}\ \frac{\abs{e}}{4}\right). 
\end{align}

\subsection{Li$_3$N} 
The third example is $\text{Li}_{3}\text{N}$ crystals, whose crytsal structure and electronic band
structure is shown in Fig.~\ref{fig:DFT}(c) belonging to the SG No.~191 ($P$6/$mmm$). 
Li nuclei are located at the WP $1b$ and $2c$ and N nuclei, each with 2 core electrons, are located at the WP $1a$.
In the DFT calculations, we only consider 14 valence electrons in the primitive cell, giving rise to 7 occupied bands in Fig.~\ref{fig:DFT}(c).
Here we assume that the crystal shape is a hexagonal prism in the hexagonal class, and its center is located at the WP $1a$. 
In this case, from Tab.~\ref{Tab:k-space_filling_anomaly}, the corner charge formula $Q^{\text{No.~191}}_{\text{corner}}$ for the SG No.~191 is given by 
\begin{equation}
    \begin{aligned}[b]
        Q^{\text{No.~191}}_{\text{corner}} =& -\frac{\abs{e}}{12}(v_{A}^{2-} + v_{A}^{4+} + v_{A}^{4-} + v_{A}^{5-} + v_{\Gamma}^{2+} \\
        &  + v_{\Gamma}^{3+}+ v_{\Gamma}^{3-} + v_{\Gamma}^{5+} + v_{L}^{1-} + v_{M}^{1+} - m_{a})
    \end{aligned}
\end{equation}
in terms of modulo $\abs{e}/6$. 
In the same way as the cases of $\text{CsLi}_{2}\text{Cl}_3$ and KN$_{3}$, we find that $v_{A}^{2-}=2$, $v_{A}^{4-}=1$, $v_{A}^{4+}=v_{A}^{5-}=v_{\Gamma}^{2+}=v_{\Gamma}^{3+}=v_{\Gamma}^{3-}=v_{\Gamma}^{5+}=v_{L}^{1-}=0$, $v_{M}^{1+}=3$, and $m_{a}=5$. 
Thus the fractional corner charge $Q^{\text{No.~191}}_{\text{corner}}$ for Li$_{3}$N crystal is 
\begin{align}
    Q^{\text{No.~191}}_{\text{corner}} = -\frac{\abs{e}}{12}\ \ (\text{mod}\ \frac{\abs{e}}{6}). 
\end{align}

\section{Conclusion \label{Sec:5}}

In this paper, we derived the formulas for quantized fractional corner charges in all the cylindrical crystal shapes with a quantized corner charge. 
Such crystal shapes are vertex-transitive polyhedra in the orthorhombic, tetragonal, hexagonal, and trigonal classes. 
Their corresponding corner charges are listed in Tabs.~\ref{tab:corner_charge_orth}, \ref{tab:corner_charge_tetra}, and \ref{tab:corner_charge_hexa}. 
We found that the corner charge formulas are universally given by $-\Delta a\abs{e} / N$ (mod $2\abs{e}/N$), ($\Delta a = n_{a} - m_{a}$, $n_{a}$: total electronic charge at WP $a$, $m_{a}$: total ionic charge at WP $a$) even if a WP at the center of a crystal is not the WP $1a$. 
This result is obtained by a lengthy calculation for the respective crystal shapes and SGs, and it is surprising that all the cases lead to the same universal result. 
We also found that the formulas of corner charge in the trigonal crystal shapes are always trivial due to possible relaxation of electronic states and ionic positions near the boundaries. 

Finally, by using the method of the EBR matrix, we derived the $k$-space formulas for the filling anomaly and corner charge for the cylindrical SGs. 
They are written in terms of the number of irreps of occupied states at HSPs in $k$ space, except for the SGs No.~89, 97, and 177 with TRS. 
For the above three SGs, we introduced a $\mathbb{Z}_{2}$ topological invariant $\xi$ for the SG No.~16 with TRS \cite{PhysRevB.109.085114,Ono=Shiozaki_top_inv}, and constructed the $k$-space formulas by incorporating this invariant $\xi$. 
The results are summarized in Tab.~\ref{Tab:k-space_filling_anomaly} for the spinless cases, while in the spinful cases, the filling anomaly is given by $\Delta a = -m_{a}$ (mod 2). 
We note that this invariant $\xi$ is valid only for systems with TRS, and thus the $k$-space formulas of the filling anomaly and corner charge without TRS are left as a future work. 
Moreover, we show that the CsLi$_{2}$Cl$_{3}$, KN$_{3}$, and Li$_{3}$N are candidate materials with a quantized corner charge by using the $ab$ $initio$ calculations. 

To summarize, this study and the previous work \cite{PhysRevB.109.085114} constitute a full list of real-space formulas of the corner charge for all the SGs and all of the corresponding crystal shapes with a quantized corner charge. 
Likewise, all the $k$-space formulas are obtained for all the SGs with TRS and for all the crystal shapes with quantized corner charges.

\begin{acknowledgments}
We thank S. Ono and K. Shiozaki for useful comments.
S. M. is supported by Japan Society for the Promotion of Science (JSPS) KAKENHI Grant No.~JP22H00108, and JP22K18687, and JP24H02231. 
\end{acknowledgments}

\appendix
\section{$\mathbb{Z}_{2}$ topological invariant for the SG No.~16 with TRS \label{Ap:inv}}
Here, we explain the $\mathbb{Z}_{2}$ topological invariant $xi$ for the SG No.~16 ($P222$) with TRS used in Sec.~\ref{Sec.4}, following Ref.~\cite{Ono=Shiozaki_top_inv}. 
The WPs and HSPs in the SG No.~16 are shown in Figs.~\ref{SGNo.16}(a) and (b), respectively. 
The little group on high-symmetry lines along the $C_{2\mu} \ (\mu = x, y,z)$ axis is $C_{2} = \{E, C_{2\mu} \}$, and that at the HSPs is $D_{2}$, whose character table are shown in Tab.~\ref{tab:D2}.
We call the irreps of $C_{2}$ with $C_{2} = +1$ and $-1$ as $A$ and $B$, respectively.
 The symmetry constraints on the Bloch Hamiltonian $H(\bm{k})$ is summarized as
\begin{align}
    &V_T(\bm{k}) H(\bm{k})^* V_T(\bm{k})^\dag = H(-\bm{k}), \\
    &V_{C_{2\mu}}(\bm{k}) H(\bm{k}) V_{C_{2\mu}}(\bm{k})^\dag = H(C_{2\mu}\bm{k}), \quad \mu=x,y,z,  \\
    &V_T(-\bm{k})V_T(\bm{k})^*=1, \\
    &V_{C_{2\mu}}(C_{2\mu}\bm{k})V_{C_{2\mu}}(\bm{k})=1, \quad \mu=x,y,z,  \\
    &V_T(C_{2\mu}\bm{k})V_{C_{2\mu}}(\bm{k})^* = V_{C_{2\mu}}(-\bm{k})V_T(\bm{k}),
\end{align}
where $V_T(\bm{k})$ is the time-reversal operator and $V_{C_{2\mu}}(\bm{k})$ is the $C_{2\mu}\ (\mu = x, y, z)$ rotational operator along the $C_{2\mu}$ axis, respectively. 
On each high-symmetry line segment $\bm{\Pi}~\bm{\Pi}^{\prime}$ connecting two HSPs $\bm{\Pi}$ and $\bm{\Pi}^{\prime}$, we introduce a Wilson line $e^{i\gamma^{B}_{\bm{\Pi}\bm{\Pi}^{\prime}}}$ of the $B$-irrep as follows. 
 Let $\Phi^{B}(\bm{k})= (\ket{u^{B}_{1}},\cdots,\ket{u^{B}_{n}})$ be an orthogonal set of occupation Bloch states with the $B$-irrep on
the line segment $\bm{\Pi}~\bm{\Pi}^{\prime}$. 
Here, $n$ is the number of the $B$-irreps.
The Wilson line $e^{i\gamma^{B}_{\bm{\Pi}\bm{\Pi}^{\prime}}}$ is defined as
\begin{align}
    \gamma^B_{\bm{\Pi}\bm{\Pi}'} = {\rm Arg} \prod_{\bm{k}=\bm{\Pi}}^{\bm{\Pi}'-\delta \bm{k}} \det \left[ \Phi^B(\bm{k} +\delta \bm{k})^\dag \Phi^B(\bm{k}) \right], 
\end{align}
where $\delta \bm{k}$ is a small displacement vector along the $\bm{\Pi}~\bm{\Pi}^{\prime}$ line. 
When $n = 0$, we set $e^{i\gamma^{B}_{\bm{\Pi}\bm{\Pi}^{\prime}}} = 1$. 
The Wilson line $e^{i\gamma^B_{\bm{\Pi}\bm{\Pi}'}}$ is not gauge invariant as it changes under the gauge transformation at the endpoints $\Phi^B(\bm{k}) \mapsto \Phi^B(\bm{k}) W^B(\bm{k})$ with $W^B(\bm{k}) \in U(n)$ a unitary matrix for $\bm{k} = \bm{\Pi}, \bm{\Pi}'$. 
We partially fix a gauge of $\Phi^B(\bm{\Pi})$ of all the HSPs $\bm{\Pi}$ as follows.
At each HSP $\bm{\Pi}$, the occupied Bloch frame is decomposed into $\Phi^{\rho}(\bm{\Pi})$ of four irreps $\rho \in \{A, B_1, B_2, B_3\}$ listed in Table~\ref{tab:D2}. 
In computing the Wilson line of the line segment parallel to the $C_{2x}$, $C_{2y}$, or $C_{2z}$-axis, the Bloch frame $\Phi^B(\bm{\Pi})$ at the endpoint $\bm{\Pi}$ is set as:
\begin{align}
\Phi^B(\bm{\Pi}) = 
\begin{cases} 
(\Phi^{B_1}(\bm{\Pi}), \Phi^{B_2}(\bm{\Pi})) & \text{for } C_{2x}\text{-axis}, \\
(\Phi^{B_1}(\bm{\Pi}), \Phi^{B_3}(\bm{\Pi})) & \text{for } C_{2y}\text{-axis}, \\
(\Phi^{B_2}(\bm{\Pi}), \Phi^{B_3}(\bm{\Pi})) & \text{for } C_{2z}\text{-axis}.
\end{cases}
\label{eq:gauge_fixed_cond_HSP}
\end{align}
Moreover, since even under Eq.~(\ref{eq:gauge_fixed_cond_HSP}), some freedom on the gauge choice remains, we incorporate the contribution from this freedom into the $\mathbb{Z}_{2}$ invariant as the term $\rho$.
Thus the $\mathbb{Z}_{2}$ invariant $\xi$ is given as \cite{PhysRevB.109.085114,Ono=Shiozaki_top_inv}:
\begin{widetext}
\begin{equation}
    \begin{aligned}[b]
        (-1)^{\xi} =& e^{i \gamma^B_{\bm{\Gamma}\mathbf{X}}}e^{i \gamma^B_{\mathbf{X}\mathbf{S}}}e^{i \gamma^B_{\mathbf{S}\mathbf{Y}}}e^{i \gamma^B_{\mathbf{Y}\bm{\Gamma}}} \times e^{i \gamma^B_{\mathbf{Z}\mathbf{T}}}e^{i \gamma^B_{\mathbf{T}\mathbf{R}}}e^{i \gamma^B_{\mathbf{R}\mathbf{U}}}e^{i \gamma^B_{\mathbf{U}\mathbf{Z}}} \times e^{i \gamma^B_{\mathbf{Z}\bm{\Gamma}}}e^{i \gamma^B_{\mathbf{X}\mathbf{U}}}e^{i \gamma^B_{\mathbf{Y}\mathbf{T}}}e^{i \gamma^B_{\mathbf{R}\mathbf{S}}} \\
        &\times \prod_{\bm{\Pi} \in \{\bm{\Gamma},\mathbf{S},\mathbf{U},\mathbf{T}\}} \text{det}[(\Phi^{B_{3}}_{\bm{\Pi}})^{\dag}V_{T}(\bm{\Pi})(\Phi^{B_{3}}_{\bm{\Pi}})^{*}]^{-1} \times \prod_{\bm{\Pi} \in \{\mathbf{X},\mathbf{Y},\mathbf{Z},\mathbf{R}\}}\text{det}[(\Phi^{B_{3}}_{\bm{\Pi}})^{\dag}V_{T}(\bm{\Pi})(\Phi^{B_{3}}_{\bm{\Pi}})^{*}] \\
        & \times \exp\left[ -\frac{1}{2}\oint_{\bm{\Gamma}\rightarrow\mathbf{X}\rightarrow\mathbf{S}\rightarrow\mathbf{Y}\rightarrow\bm{\Gamma}} \text{Tr}[(\Phi^{B}(\bm{k}))^{\dag}V_{C_{2z}T}(\bm{k})d V_{C_{2z}T}(\bm{k})^{\dag} \Phi^{B}(\bm{k})] \right] \\
        & \times \exp\left[ \frac{1}{2}\oint_{\mathbf{Z}\rightarrow\mathbf{U}\rightarrow\mathbf{R}\rightarrow\mathbf{T}\rightarrow\mathbf{Z}} \text{Tr}[(\Phi^{B}(\bm{k}))^{\dag}V_{C_{2z}T}(\bm{k})d V_{C_{2z}T}(\bm{k})^{\dag} \Phi^{B}(\bm{k})] \right] \\
        & \times \exp\left[ -\frac{1}{2}(\int_{\mathbf{Z}\rightarrow\bm{\Gamma}} + \int_{\mathbf{X}\rightarrow\mathbf{U}} + \int_{\mathbf{Y}\rightarrow\mathbf{T}} + \int_{\mathbf{R}\rightarrow\mathbf{S}})\text{Tr}[(\Phi^{B}(\bm{k}))^{\dag}V_{C_{2x}T}(\bm{k})d V_{C_{2x}T}(\bm{k})^{\dag} \Phi^{B}(\bm{k})] \right] \times (-1)^{\rho},
        \label{eq:new_invariant}
    \end{aligned}
\end{equation}
Here, $V_{C_{2\mu}T}(\bm{k}) = V_{C_{2\mu}}(-\bm{k})V_T(\bm{k})$, and $\rho \in \{0,1\}$ is a quadratic function determined by the irreps at the HSPs, 
\begin{align}
\rho 
&:= n_{R}^{B_3} \left(n_{S}^{B_3}+n_{T}^{B_2}+n_{T}^{B_3}+n_{U}^{B_2}+n_{U}^{B_3}+n_{Z}^{B_3}\right)
+n_{R}^{B_2} \left(n_{T}^{B_3}+n_{U}^{B_3}+n_{X}^{B_3}+n_{Y}^{B_3}\right) \nonumber \\
&+n_{T}^{B_2} \left(n_{S}^{B_3}+n_{U}^{B_3}+n_{X}^{B_3}\right)+n_{Y}^{B_3} \left(n_{S}^{B_3}+n_{U}^{B_2}+n_{X}^{B_3}+n_{Z}^{B_3}\right)
+n_{S}^{B_3} n_{U}^{B_2}+n_{X}^{B_3} \left(n_{S}^{B_3}+n_{Z}^{B_3}\right)+n_{S}^{B_3} n_{Z}^{B_3}\nonumber \\
&+n_{T}^{B_3} \left(n_{U}^{B_2}+n_{U}^{B_3}+n_{Y}^{B_3}+n_{Z}^{B_3}\right)+n_{U}^{B_3} \left(n_{X}^{B_3}+n_{Z}^{B_3}\right) + n_{S}^{B_3}+n_{T}^{B_3}+n_{U}^{B_3} \quad \mod 2.
\end{align}
\end{widetext}

\bibliography{reference}

\begin{thebibliography}{56}%
\makeatletter
\providecommand \@ifxundefined [1]{%
 \@ifx{#1\undefined}
}%
\providecommand \@ifnum [1]{%
 \ifnum #1\expandafter \@firstoftwo
 \else \expandafter \@secondoftwo
 \fi
}%
\providecommand \@ifx [1]{%
 \ifx #1\expandafter \@firstoftwo
 \else \expandafter \@secondoftwo
 \fi
}%
\providecommand \natexlab [1]{#1}%
\providecommand \enquote  [1]{``#1''}%
\providecommand \bibnamefont  [1]{#1}%
\providecommand \bibfnamefont [1]{#1}%
\providecommand \citenamefont [1]{#1}%
\providecommand \href@noop [0]{\@secondoftwo}%
\providecommand \href [0]{\begingroup \@sanitize@url \@href}%
\providecommand \@href[1]{\@@startlink{#1}\@@href}%
\providecommand \@@href[1]{\endgroup#1\@@endlink}%
\providecommand \@sanitize@url [0]{\catcode `\\12\catcode `\$12\catcode `\&12\catcode `\#12\catcode `\^12\catcode `\_12\catcode `\%12\relax}%
\providecommand \@@startlink[1]{}%
\providecommand \@@endlink[0]{}%
\providecommand \url  [0]{\begingroup\@sanitize@url \@url }%
\providecommand \@url [1]{\endgroup\@href {#1}{\urlprefix }}%
\providecommand \urlprefix  [0]{URL }%
\providecommand \Eprint [0]{\href }%
\providecommand \doibase [0]{https://doi.org/}%
\providecommand \selectlanguage [0]{\@gobble}%
\providecommand \bibinfo  [0]{\@secondoftwo}%
\providecommand \bibfield  [0]{\@secondoftwo}%
\providecommand \translation [1]{[#1]}%
\providecommand \BibitemOpen [0]{}%
\providecommand \bibitemStop [0]{}%
\providecommand \bibitemNoStop [0]{.\EOS\space}%
\providecommand \EOS [0]{\spacefactor3000\relax}%
\providecommand \BibitemShut  [1]{\csname bibitem#1\endcsname}%
\let\auto@bib@innerbib\@empty
\bibitem [{\citenamefont {Fu}\ and\ \citenamefont {Kane}(2007)}]{PhysRevB.76.045302}%
  \BibitemOpen
  \bibfield  {author} {\bibinfo {author} {\bibfnamefont {L.}~\bibnamefont {Fu}}\ and\ \bibinfo {author} {\bibfnamefont {C.~L.}\ \bibnamefont {Kane}},\ }\bibfield  {title} {\bibinfo {title} {Topological insulators with inversion symmetry},\ }\href {https://doi.org/10.1103/PhysRevB.76.045302} {\bibfield  {journal} {\bibinfo  {journal} {Phys. Rev. B}\ }\textbf {\bibinfo {volume} {76}},\ \bibinfo {pages} {045302} (\bibinfo {year} {2007})}\BibitemShut {NoStop}%
\bibitem [{\citenamefont {Thouless}(1983)}]{PhysRevB.27.6083}%
  \BibitemOpen
  \bibfield  {author} {\bibinfo {author} {\bibfnamefont {D.~J.}\ \bibnamefont {Thouless}},\ }\bibfield  {title} {\bibinfo {title} {Quantization of particle transport},\ }\href {https://doi.org/10.1103/PhysRevB.27.6083} {\bibfield  {journal} {\bibinfo  {journal} {Phys. Rev. B}\ }\textbf {\bibinfo {volume} {27}},\ \bibinfo {pages} {6083} (\bibinfo {year} {1983})}\BibitemShut {NoStop}%
\bibitem [{\citenamefont {Fu}\ \emph {et~al.}(2007)\citenamefont {Fu}, \citenamefont {Kane},\ and\ \citenamefont {Mele}}]{PhysRevLett.98.106803}%
  \BibitemOpen
  \bibfield  {author} {\bibinfo {author} {\bibfnamefont {L.}~\bibnamefont {Fu}}, \bibinfo {author} {\bibfnamefont {C.~L.}\ \bibnamefont {Kane}},\ and\ \bibinfo {author} {\bibfnamefont {E.~J.}\ \bibnamefont {Mele}},\ }\bibfield  {title} {\bibinfo {title} {Topological insulators in three dimensions},\ }\href {https://doi.org/10.1103/PhysRevLett.98.106803} {\bibfield  {journal} {\bibinfo  {journal} {Phys. Rev. Lett.}\ }\textbf {\bibinfo {volume} {98}},\ \bibinfo {pages} {106803} (\bibinfo {year} {2007})}\BibitemShut {NoStop}%
\bibitem [{\citenamefont {Kane}\ and\ \citenamefont {Mele}(2005{\natexlab{a}})}]{PhysRevLett.95.146802}%
  \BibitemOpen
  \bibfield  {author} {\bibinfo {author} {\bibfnamefont {C.~L.}\ \bibnamefont {Kane}}\ and\ \bibinfo {author} {\bibfnamefont {E.~J.}\ \bibnamefont {Mele}},\ }\bibfield  {title} {\bibinfo {title} {${Z}_{2}$ topological order and the quantum spin {H}all effect},\ }\href {https://doi.org/10.1103/PhysRevLett.95.146802} {\bibfield  {journal} {\bibinfo  {journal} {Phys. Rev. Lett.}\ }\textbf {\bibinfo {volume} {95}},\ \bibinfo {pages} {146802} (\bibinfo {year} {2005}{\natexlab{a}})}\BibitemShut {NoStop}%
\bibitem [{\citenamefont {Kane}\ and\ \citenamefont {Mele}(2005{\natexlab{b}})}]{PhysRevLett.95.226801}%
  \BibitemOpen
  \bibfield  {author} {\bibinfo {author} {\bibfnamefont {C.~L.}\ \bibnamefont {Kane}}\ and\ \bibinfo {author} {\bibfnamefont {E.~J.}\ \bibnamefont {Mele}},\ }\bibfield  {title} {\bibinfo {title} {Quantum spin {H}all effect in graphene},\ }\href {https://doi.org/10.1103/PhysRevLett.95.226801} {\bibfield  {journal} {\bibinfo  {journal} {Phys. Rev. Lett.}\ }\textbf {\bibinfo {volume} {95}},\ \bibinfo {pages} {226801} (\bibinfo {year} {2005}{\natexlab{b}})}\BibitemShut {NoStop}%
\bibitem [{\citenamefont {Fu}\ and\ \citenamefont {Kane}(2006)}]{PhysRevB.74.195312}%
  \BibitemOpen
  \bibfield  {author} {\bibinfo {author} {\bibfnamefont {L.}~\bibnamefont {Fu}}\ and\ \bibinfo {author} {\bibfnamefont {C.~L.}\ \bibnamefont {Kane}},\ }\bibfield  {title} {\bibinfo {title} {Time reversal polarization and a ${Z}_{2}$ adiabatic spin pump},\ }\href {https://doi.org/10.1103/PhysRevB.74.195312} {\bibfield  {journal} {\bibinfo  {journal} {Phys. Rev. B}\ }\textbf {\bibinfo {volume} {74}},\ \bibinfo {pages} {195312} (\bibinfo {year} {2006})}\BibitemShut {NoStop}%
\bibitem [{\citenamefont {König}\ \emph {et~al.}(2007)\citenamefont {König}, \citenamefont {Wiedmann}, \citenamefont {Brüne}, \citenamefont {Roth}, \citenamefont {Buhmann}, \citenamefont {Molenkamp}, \citenamefont {Qi},\ and\ \citenamefont {Zhang}}]{doi:10.1126/science.1148047}%
  \BibitemOpen
  \bibfield  {author} {\bibinfo {author} {\bibfnamefont {M.}~\bibnamefont {König}}, \bibinfo {author} {\bibfnamefont {S.}~\bibnamefont {Wiedmann}}, \bibinfo {author} {\bibfnamefont {C.}~\bibnamefont {Brüne}}, \bibinfo {author} {\bibfnamefont {A.}~\bibnamefont {Roth}}, \bibinfo {author} {\bibfnamefont {H.}~\bibnamefont {Buhmann}}, \bibinfo {author} {\bibfnamefont {L.~W.}\ \bibnamefont {Molenkamp}}, \bibinfo {author} {\bibfnamefont {X.-L.}\ \bibnamefont {Qi}},\ and\ \bibinfo {author} {\bibfnamefont {S.-C.}\ \bibnamefont {Zhang}},\ }\bibfield  {title} {\bibinfo {title} {Quantum spin {H}all insulator state in {H}g{T}e quantum wells},\ }\href {https://doi.org/10.1126/science.1148047} {\bibfield  {journal} {\bibinfo  {journal} {Science}\ }\textbf {\bibinfo {volume} {318}},\ \bibinfo {pages} {766} (\bibinfo {year} {2007})}\BibitemShut {NoStop}%
\bibitem [{\citenamefont {Bernevig}\ \emph {et~al.}(2006)\citenamefont {Bernevig}, \citenamefont {Hughes},\ and\ \citenamefont {Zhang}}]{doi:10.1126/science.1133734}%
  \BibitemOpen
  \bibfield  {author} {\bibinfo {author} {\bibfnamefont {B.~A.}\ \bibnamefont {Bernevig}}, \bibinfo {author} {\bibfnamefont {T.~L.}\ \bibnamefont {Hughes}},\ and\ \bibinfo {author} {\bibfnamefont {S.-C.}\ \bibnamefont {Zhang}},\ }\bibfield  {title} {\bibinfo {title} {Quantum spin {H}all effect and topological phase transition in {H}g{T}e quantum wells},\ }\href {https://doi.org/10.1126/science.1133734} {\bibfield  {journal} {\bibinfo  {journal} {Science}\ }\textbf {\bibinfo {volume} {314}},\ \bibinfo {pages} {1757} (\bibinfo {year} {2006})}\BibitemShut {NoStop}%
\bibitem [{\citenamefont {Teo}\ \emph {et~al.}(2008)\citenamefont {Teo}, \citenamefont {Fu},\ and\ \citenamefont {Kane}}]{PhysRevB.78.045426}%
  \BibitemOpen
  \bibfield  {author} {\bibinfo {author} {\bibfnamefont {J.~C.~Y.}\ \bibnamefont {Teo}}, \bibinfo {author} {\bibfnamefont {L.}~\bibnamefont {Fu}},\ and\ \bibinfo {author} {\bibfnamefont {C.~L.}\ \bibnamefont {Kane}},\ }\bibfield  {title} {\bibinfo {title} {Surface states and topological invariants in three-dimensional topological insulators: Application to ${{\text{Bi}}}_{1\ensuremath{-}x}{{\text{Sb}}}_{x}$},\ }\href {https://doi.org/10.1103/PhysRevB.78.045426} {\bibfield  {journal} {\bibinfo  {journal} {Phys. Rev. B}\ }\textbf {\bibinfo {volume} {78}},\ \bibinfo {pages} {045426} (\bibinfo {year} {2008})}\BibitemShut {NoStop}%
\bibitem [{\citenamefont {Fu}(2011)}]{PhysRevLett.106.106802}%
  \BibitemOpen
  \bibfield  {author} {\bibinfo {author} {\bibfnamefont {L.}~\bibnamefont {Fu}},\ }\bibfield  {title} {\bibinfo {title} {Topological crystalline insulators},\ }\href {https://doi.org/10.1103/PhysRevLett.106.106802} {\bibfield  {journal} {\bibinfo  {journal} {Phys. Rev. Lett.}\ }\textbf {\bibinfo {volume} {106}},\ \bibinfo {pages} {106802} (\bibinfo {year} {2011})}\BibitemShut {NoStop}%
\bibitem [{\citenamefont {Benalcazar}\ \emph {et~al.}(2017)\citenamefont {Benalcazar}, \citenamefont {Bernevig},\ and\ \citenamefont {Hughes}}]{PhysRevB.96.245115}%
  \BibitemOpen
  \bibfield  {author} {\bibinfo {author} {\bibfnamefont {W.~A.}\ \bibnamefont {Benalcazar}}, \bibinfo {author} {\bibfnamefont {B.~A.}\ \bibnamefont {Bernevig}},\ and\ \bibinfo {author} {\bibfnamefont {T.~L.}\ \bibnamefont {Hughes}},\ }\bibfield  {title} {\bibinfo {title} {Electric multipole moments, topological multipole moment pumping, and chiral hinge states in crystalline insulators},\ }\href {https://doi.org/10.1103/PhysRevB.96.245115} {\bibfield  {journal} {\bibinfo  {journal} {Phys. Rev. B}\ }\textbf {\bibinfo {volume} {96}},\ \bibinfo {pages} {245115} (\bibinfo {year} {2017})}\BibitemShut {NoStop}%
\bibitem [{\citenamefont {Fang}\ and\ \citenamefont {Fu}(2015)}]{PhysRevB.91.161105}%
  \BibitemOpen
  \bibfield  {author} {\bibinfo {author} {\bibfnamefont {C.}~\bibnamefont {Fang}}\ and\ \bibinfo {author} {\bibfnamefont {L.}~\bibnamefont {Fu}},\ }\bibfield  {title} {\bibinfo {title} {New classes of three-dimensional topological crystalline insulators: Nonsymmorphic and magnetic},\ }\href {https://doi.org/10.1103/PhysRevB.91.161105} {\bibfield  {journal} {\bibinfo  {journal} {Phys. Rev. B}\ }\textbf {\bibinfo {volume} {91}},\ \bibinfo {pages} {161105} (\bibinfo {year} {2015})}\BibitemShut {NoStop}%
\bibitem [{\citenamefont {Watanabe}\ and\ \citenamefont {Fu}(2017)}]{PhysRevB.95.081107}%
  \BibitemOpen
  \bibfield  {author} {\bibinfo {author} {\bibfnamefont {H.}~\bibnamefont {Watanabe}}\ and\ \bibinfo {author} {\bibfnamefont {L.}~\bibnamefont {Fu}},\ }\bibfield  {title} {\bibinfo {title} {Topological crystalline magnets: Symmetry-protected topological phases of fermions},\ }\href {https://doi.org/10.1103/PhysRevB.95.081107} {\bibfield  {journal} {\bibinfo  {journal} {Phys. Rev. B}\ }\textbf {\bibinfo {volume} {95}},\ \bibinfo {pages} {081107} (\bibinfo {year} {2017})}\BibitemShut {NoStop}%
\bibitem [{\citenamefont {Shiozaki}\ and\ \citenamefont {Sato}(2014)}]{PhysRevB.90.165114}%
  \BibitemOpen
  \bibfield  {author} {\bibinfo {author} {\bibfnamefont {K.}~\bibnamefont {Shiozaki}}\ and\ \bibinfo {author} {\bibfnamefont {M.}~\bibnamefont {Sato}},\ }\bibfield  {title} {\bibinfo {title} {Topology of crystalline insulators and superconductors},\ }\href {https://doi.org/10.1103/PhysRevB.90.165114} {\bibfield  {journal} {\bibinfo  {journal} {Phys. Rev. B}\ }\textbf {\bibinfo {volume} {90}},\ \bibinfo {pages} {165114} (\bibinfo {year} {2014})}\BibitemShut {NoStop}%
\bibitem [{\citenamefont {Kruthoff}\ \emph {et~al.}(2017)\citenamefont {Kruthoff}, \citenamefont {de~Boer}, \citenamefont {van Wezel}, \citenamefont {Kane},\ and\ \citenamefont {Slager}}]{PhysRevX.7.041069}%
  \BibitemOpen
  \bibfield  {author} {\bibinfo {author} {\bibfnamefont {J.}~\bibnamefont {Kruthoff}}, \bibinfo {author} {\bibfnamefont {J.}~\bibnamefont {de~Boer}}, \bibinfo {author} {\bibfnamefont {J.}~\bibnamefont {van Wezel}}, \bibinfo {author} {\bibfnamefont {C.~L.}\ \bibnamefont {Kane}},\ and\ \bibinfo {author} {\bibfnamefont {R.-J.}\ \bibnamefont {Slager}},\ }\bibfield  {title} {\bibinfo {title} {{Topological Classification of Crystalline Insulators through Band Structure Combinatorics}},\ }\href {https://doi.org/10.1103/PhysRevX.7.041069} {\bibfield  {journal} {\bibinfo  {journal} {Phys. Rev. X}\ }\textbf {\bibinfo {volume} {7}},\ \bibinfo {pages} {041069} (\bibinfo {year} {2017})}\BibitemShut {NoStop}%
\bibitem [{\citenamefont {van Miert}\ and\ \citenamefont {Ortix}(2018)}]{PhysRevB.98.081110}%
  \BibitemOpen
  \bibfield  {author} {\bibinfo {author} {\bibfnamefont {G.}~\bibnamefont {van Miert}}\ and\ \bibinfo {author} {\bibfnamefont {C.}~\bibnamefont {Ortix}},\ }\bibfield  {title} {\bibinfo {title} {Higher-order topological insulators protected by inversion and rotoinversion symmetries},\ }\href {https://doi.org/10.1103/PhysRevB.98.081110} {\bibfield  {journal} {\bibinfo  {journal} {Phys. Rev. B}\ }\textbf {\bibinfo {volume} {98}},\ \bibinfo {pages} {081110} (\bibinfo {year} {2018})}\BibitemShut {NoStop}%
\bibitem [{\citenamefont {Song}\ \emph {et~al.}(2017)\citenamefont {Song}, \citenamefont {Fang},\ and\ \citenamefont {Fang}}]{PhysRevLett.119.246402}%
  \BibitemOpen
  \bibfield  {author} {\bibinfo {author} {\bibfnamefont {Z.}~\bibnamefont {Song}}, \bibinfo {author} {\bibfnamefont {Z.}~\bibnamefont {Fang}},\ and\ \bibinfo {author} {\bibfnamefont {C.}~\bibnamefont {Fang}},\ }\bibfield  {title} {\bibinfo {title} {$(d\ensuremath{-}2)$-dimensional edge states of rotation symmetry protected topological states},\ }\href {https://doi.org/10.1103/PhysRevLett.119.246402} {\bibfield  {journal} {\bibinfo  {journal} {Phys. Rev. Lett.}\ }\textbf {\bibinfo {volume} {119}},\ \bibinfo {pages} {246402} (\bibinfo {year} {2017})}\BibitemShut {NoStop}%
\bibitem [{\citenamefont {Schindler}\ \emph {et~al.}(2018{\natexlab{a}})\citenamefont {Schindler}, \citenamefont {Wang}, \citenamefont {Vergniory}, \citenamefont {Cook}, \citenamefont {Murani}, \citenamefont {Sengupta}, \citenamefont {Kasumov}, \citenamefont {Deblock}, \citenamefont {Jeon}, \citenamefont {Drozdov}, \citenamefont {Bouchiat}, \citenamefont {Gu{\'{e}}ron}, \citenamefont {Yazdani}, \citenamefont {Bernevig},\ and\ \citenamefont {Neupert}}]{Schindler_2018}%
  \BibitemOpen
  \bibfield  {author} {\bibinfo {author} {\bibfnamefont {F.}~\bibnamefont {Schindler}}, \bibinfo {author} {\bibfnamefont {Z.}~\bibnamefont {Wang}}, \bibinfo {author} {\bibfnamefont {M.~G.}\ \bibnamefont {Vergniory}}, \bibinfo {author} {\bibfnamefont {A.~M.}\ \bibnamefont {Cook}}, \bibinfo {author} {\bibfnamefont {A.}~\bibnamefont {Murani}}, \bibinfo {author} {\bibfnamefont {S.}~\bibnamefont {Sengupta}}, \bibinfo {author} {\bibfnamefont {A.~Y.}\ \bibnamefont {Kasumov}}, \bibinfo {author} {\bibfnamefont {R.}~\bibnamefont {Deblock}}, \bibinfo {author} {\bibfnamefont {S.}~\bibnamefont {Jeon}}, \bibinfo {author} {\bibfnamefont {I.}~\bibnamefont {Drozdov}}, \bibinfo {author} {\bibfnamefont {H.}~\bibnamefont {Bouchiat}}, \bibinfo {author} {\bibfnamefont {S.}~\bibnamefont {Gu{\'{e}}ron}}, \bibinfo {author} {\bibfnamefont {A.}~\bibnamefont {Yazdani}}, \bibinfo {author} {\bibfnamefont {B.~A.}\ \bibnamefont {Bernevig}},\ and\ \bibinfo {author} {\bibfnamefont {T.}~\bibnamefont {Neupert}},\ }\bibfield  {title} {\bibinfo
  {title} {Higher-order topology in bismuth},\ }\href {https://doi.org/10.1038/s41567-018-0224-7} {\bibfield  {journal} {\bibinfo  {journal} {Nature Physics}\ }\textbf {\bibinfo {volume} {14}},\ \bibinfo {pages} {918} (\bibinfo {year} {2018}{\natexlab{a}})}\BibitemShut {NoStop}%
\bibitem [{\citenamefont {Schindler}\ \emph {et~al.}(2018{\natexlab{b}})\citenamefont {Schindler}, \citenamefont {Cook}, \citenamefont {Vergniory}, \citenamefont {Wang}, \citenamefont {Parkin}, \citenamefont {Bernevig},\ and\ \citenamefont {Neupert}}]{doi:10.1126/sciadv.aat0346}%
  \BibitemOpen
  \bibfield  {author} {\bibinfo {author} {\bibfnamefont {F.}~\bibnamefont {Schindler}}, \bibinfo {author} {\bibfnamefont {A.~M.}\ \bibnamefont {Cook}}, \bibinfo {author} {\bibfnamefont {M.~G.}\ \bibnamefont {Vergniory}}, \bibinfo {author} {\bibfnamefont {Z.}~\bibnamefont {Wang}}, \bibinfo {author} {\bibfnamefont {S.~S.~P.}\ \bibnamefont {Parkin}}, \bibinfo {author} {\bibfnamefont {B.~A.}\ \bibnamefont {Bernevig}},\ and\ \bibinfo {author} {\bibfnamefont {T.}~\bibnamefont {Neupert}},\ }\bibfield  {title} {\bibinfo {title} {Higher-order topological insulators},\ }\href {https://doi.org/10.1126/sciadv.aat0346} {\bibfield  {journal} {\bibinfo  {journal} {Science Advances}\ }\textbf {\bibinfo {volume} {4}},\ \bibinfo {pages} {eaat0346} (\bibinfo {year} {2018}{\natexlab{b}})}\BibitemShut {NoStop}%
\bibitem [{\citenamefont {Agarwala}\ \emph {et~al.}(2020)\citenamefont {Agarwala}, \citenamefont {Juri\ifmmode \check{c}\else \v{c}\fi{}i\ifmmode~\acute{c}\else \'{c}\fi{}},\ and\ \citenamefont {Roy}}]{PhysRevResearch.2.012067}%
  \BibitemOpen
  \bibfield  {author} {\bibinfo {author} {\bibfnamefont {A.}~\bibnamefont {Agarwala}}, \bibinfo {author} {\bibfnamefont {V.}~\bibnamefont {Juri\ifmmode \check{c}\else \v{c}\fi{}i\ifmmode~\acute{c}\else \'{c}\fi{}}},\ and\ \bibinfo {author} {\bibfnamefont {B.}~\bibnamefont {Roy}},\ }\bibfield  {title} {\bibinfo {title} {Higher-order topological insulators in amorphous solids},\ }\href {https://doi.org/10.1103/PhysRevResearch.2.012067} {\bibfield  {journal} {\bibinfo  {journal} {Phys. Rev. Res.}\ }\textbf {\bibinfo {volume} {2}},\ \bibinfo {pages} {012067} (\bibinfo {year} {2020})}\BibitemShut {NoStop}%
\bibitem [{\citenamefont {Chen}\ \emph {et~al.}(2020)\citenamefont {Chen}, \citenamefont {Chen}, \citenamefont {Gao}, \citenamefont {Zhou},\ and\ \citenamefont {Xu}}]{PhysRevLett.124.036803}%
  \BibitemOpen
  \bibfield  {author} {\bibinfo {author} {\bibfnamefont {R.}~\bibnamefont {Chen}}, \bibinfo {author} {\bibfnamefont {C.-Z.}\ \bibnamefont {Chen}}, \bibinfo {author} {\bibfnamefont {J.-H.}\ \bibnamefont {Gao}}, \bibinfo {author} {\bibfnamefont {B.}~\bibnamefont {Zhou}},\ and\ \bibinfo {author} {\bibfnamefont {D.-H.}\ \bibnamefont {Xu}},\ }\bibfield  {title} {\bibinfo {title} {Higher-order topological insulators in quasicrystals},\ }\href {https://doi.org/10.1103/PhysRevLett.124.036803} {\bibfield  {journal} {\bibinfo  {journal} {Phys. Rev. Lett.}\ }\textbf {\bibinfo {volume} {124}},\ \bibinfo {pages} {036803} (\bibinfo {year} {2020})}\BibitemShut {NoStop}%
\bibitem [{\citenamefont {Hirayama}\ \emph {et~al.}(2020)\citenamefont {Hirayama}, \citenamefont {Takahashi}, \citenamefont {Matsuishi}, \citenamefont {Hosono},\ and\ \citenamefont {Murakami}}]{PhysRevResearch.2.043131}%
  \BibitemOpen
  \bibfield  {author} {\bibinfo {author} {\bibfnamefont {M.}~\bibnamefont {Hirayama}}, \bibinfo {author} {\bibfnamefont {R.}~\bibnamefont {Takahashi}}, \bibinfo {author} {\bibfnamefont {S.}~\bibnamefont {Matsuishi}}, \bibinfo {author} {\bibfnamefont {H.}~\bibnamefont {Hosono}},\ and\ \bibinfo {author} {\bibfnamefont {S.}~\bibnamefont {Murakami}},\ }\bibfield  {title} {\bibinfo {title} {Higher-order topological crystalline insulating phase and quantized hinge charge in topological electride apatite},\ }\href {https://doi.org/10.1103/PhysRevResearch.2.043131} {\bibfield  {journal} {\bibinfo  {journal} {Phys. Rev. Res.}\ }\textbf {\bibinfo {volume} {2}},\ \bibinfo {pages} {043131} (\bibinfo {year} {2020})}\BibitemShut {NoStop}%
\bibitem [{\citenamefont {Watanabe}\ and\ \citenamefont {Po}(2021)}]{PhysRevX.11.041064}%
  \BibitemOpen
  \bibfield  {author} {\bibinfo {author} {\bibfnamefont {H.}~\bibnamefont {Watanabe}}\ and\ \bibinfo {author} {\bibfnamefont {H.~C.}\ \bibnamefont {Po}},\ }\bibfield  {title} {\bibinfo {title} {Fractional corner charge of sodium chloride},\ }\href {https://doi.org/10.1103/PhysRevX.11.041064} {\bibfield  {journal} {\bibinfo  {journal} {Phys. Rev. X}\ }\textbf {\bibinfo {volume} {11}},\ \bibinfo {pages} {041064} (\bibinfo {year} {2021})}\BibitemShut {NoStop}%
\bibitem [{\citenamefont {Naito}\ \emph {et~al.}(2022)\citenamefont {Naito}, \citenamefont {Takahashi}, \citenamefont {Watanabe},\ and\ \citenamefont {Murakami}}]{PhysRevB.105.045126}%
  \BibitemOpen
  \bibfield  {author} {\bibinfo {author} {\bibfnamefont {K.}~\bibnamefont {Naito}}, \bibinfo {author} {\bibfnamefont {R.}~\bibnamefont {Takahashi}}, \bibinfo {author} {\bibfnamefont {H.}~\bibnamefont {Watanabe}},\ and\ \bibinfo {author} {\bibfnamefont {S.}~\bibnamefont {Murakami}},\ }\bibfield  {title} {\bibinfo {title} {Fractional hinge and corner charges in various crystal shapes with cubic symmetry},\ }\href {https://doi.org/10.1103/PhysRevB.105.045126} {\bibfield  {journal} {\bibinfo  {journal} {Phys. Rev. B}\ }\textbf {\bibinfo {volume} {105}},\ \bibinfo {pages} {045126} (\bibinfo {year} {2022})}\BibitemShut {NoStop}%
\bibitem [{\citenamefont {Wada}\ \emph {et~al.}(2024)\citenamefont {Wada}, \citenamefont {Naito}, \citenamefont {Ono}, \citenamefont {Shiozaki},\ and\ \citenamefont {Murakami}}]{PhysRevB.109.085114}%
  \BibitemOpen
  \bibfield  {author} {\bibinfo {author} {\bibfnamefont {H.}~\bibnamefont {Wada}}, \bibinfo {author} {\bibfnamefont {K.}~\bibnamefont {Naito}}, \bibinfo {author} {\bibfnamefont {S.}~\bibnamefont {Ono}}, \bibinfo {author} {\bibfnamefont {K.}~\bibnamefont {Shiozaki}},\ and\ \bibinfo {author} {\bibfnamefont {S.}~\bibnamefont {Murakami}},\ }\bibfield  {title} {\bibinfo {title} {General corner charge formulas in various tetrahedral and cubic space groups},\ }\href {https://doi.org/10.1103/PhysRevB.109.085114} {\bibfield  {journal} {\bibinfo  {journal} {Phys. Rev. B}\ }\textbf {\bibinfo {volume} {109}},\ \bibinfo {pages} {085114} (\bibinfo {year} {2024})}\BibitemShut {NoStop}%
\bibitem [{\citenamefont {Benalcazar}\ \emph {et~al.}(2019)\citenamefont {Benalcazar}, \citenamefont {Li},\ and\ \citenamefont {Hughes}}]{PhysRevB.99.245151}%
  \BibitemOpen
  \bibfield  {author} {\bibinfo {author} {\bibfnamefont {W.~A.}\ \bibnamefont {Benalcazar}}, \bibinfo {author} {\bibfnamefont {T.}~\bibnamefont {Li}},\ and\ \bibinfo {author} {\bibfnamefont {T.~L.}\ \bibnamefont {Hughes}},\ }\bibfield  {title} {\bibinfo {title} {Quantization of fractional corner charge in ${C}_{n}$-symmetric higher-order topological crystalline insulators},\ }\href {https://doi.org/10.1103/PhysRevB.99.245151} {\bibfield  {journal} {\bibinfo  {journal} {Phys. Rev. B}\ }\textbf {\bibinfo {volume} {99}},\ \bibinfo {pages} {245151} (\bibinfo {year} {2019})}\BibitemShut {NoStop}%
\bibitem [{\citenamefont {Takahashi}\ \emph {et~al.}(2021)\citenamefont {Takahashi}, \citenamefont {Zhang},\ and\ \citenamefont {Murakami}}]{PhysRevB.103.205123}%
  \BibitemOpen
  \bibfield  {author} {\bibinfo {author} {\bibfnamefont {R.}~\bibnamefont {Takahashi}}, \bibinfo {author} {\bibfnamefont {T.}~\bibnamefont {Zhang}},\ and\ \bibinfo {author} {\bibfnamefont {S.}~\bibnamefont {Murakami}},\ }\bibfield  {title} {\bibinfo {title} {General corner charge formula in two-dimensional ${C}_{n}$-symmetric higher-order topological insulators},\ }\href {https://doi.org/10.1103/PhysRevB.103.205123} {\bibfield  {journal} {\bibinfo  {journal} {Phys. Rev. B}\ }\textbf {\bibinfo {volume} {103}},\ \bibinfo {pages} {205123} (\bibinfo {year} {2021})}\BibitemShut {NoStop}%
\bibitem [{\citenamefont {Schindler}\ \emph {et~al.}(2019)\citenamefont {Schindler}, \citenamefont {Brzezi\ifmmode~\acute{n}\else \'{n}\fi{}ska}, \citenamefont {Benalcazar}, \citenamefont {Iraola}, \citenamefont {Bouhon}, \citenamefont {Tsirkin}, \citenamefont {Vergniory},\ and\ \citenamefont {Neupert}}]{PhysRevResearch.1.033074}%
  \BibitemOpen
  \bibfield  {author} {\bibinfo {author} {\bibfnamefont {F.}~\bibnamefont {Schindler}}, \bibinfo {author} {\bibfnamefont {M.}~\bibnamefont {Brzezi\ifmmode~\acute{n}\else \'{n}\fi{}ska}}, \bibinfo {author} {\bibfnamefont {W.~A.}\ \bibnamefont {Benalcazar}}, \bibinfo {author} {\bibfnamefont {M.}~\bibnamefont {Iraola}}, \bibinfo {author} {\bibfnamefont {A.}~\bibnamefont {Bouhon}}, \bibinfo {author} {\bibfnamefont {S.~S.}\ \bibnamefont {Tsirkin}}, \bibinfo {author} {\bibfnamefont {M.~G.}\ \bibnamefont {Vergniory}},\ and\ \bibinfo {author} {\bibfnamefont {T.}~\bibnamefont {Neupert}},\ }\bibfield  {title} {\bibinfo {title} {Fractional corner charges in spin-orbit coupled crystals},\ }\href {https://doi.org/10.1103/PhysRevResearch.1.033074} {\bibfield  {journal} {\bibinfo  {journal} {Phys. Rev. Res.}\ }\textbf {\bibinfo {volume} {1}},\ \bibinfo {pages} {033074} (\bibinfo {year} {2019})}\BibitemShut {NoStop}%
\bibitem [{\citenamefont {Watanabe}\ and\ \citenamefont {Ono}(2020)}]{PhysRevB.102.165120}%
  \BibitemOpen
  \bibfield  {author} {\bibinfo {author} {\bibfnamefont {H.}~\bibnamefont {Watanabe}}\ and\ \bibinfo {author} {\bibfnamefont {S.}~\bibnamefont {Ono}},\ }\bibfield  {title} {\bibinfo {title} {Corner charge and bulk multipole moment in periodic systems},\ }\href {https://doi.org/10.1103/PhysRevB.102.165120} {\bibfield  {journal} {\bibinfo  {journal} {Phys. Rev. B}\ }\textbf {\bibinfo {volume} {102}},\ \bibinfo {pages} {165120} (\bibinfo {year} {2020})}\BibitemShut {NoStop}%
\bibitem [{\citenamefont {Po}\ \emph {et~al.}(2017)\citenamefont {Po}, \citenamefont {Vishwanath},\ and\ \citenamefont {Watanabe}}]{Po_2017}%
  \BibitemOpen
  \bibfield  {author} {\bibinfo {author} {\bibfnamefont {H.~C.}\ \bibnamefont {Po}}, \bibinfo {author} {\bibfnamefont {A.}~\bibnamefont {Vishwanath}},\ and\ \bibinfo {author} {\bibfnamefont {H.}~\bibnamefont {Watanabe}},\ }\bibfield  {title} {\bibinfo {title} {Symmetry-based indicators of band topology in the 230 space groups},\ }\bibfield  {journal} {\bibinfo  {journal} {Nature Communications}\ }\textbf {\bibinfo {volume} {8, 50}},\ \href@noop {} {} (\bibinfo {year} {2017})\BibitemShut {NoStop}%
\bibitem [{\citenamefont {Watanabe}\ \emph {et~al.}(2018)\citenamefont {Watanabe}, \citenamefont {Po},\ and\ \citenamefont {Vishwanath}}]{SI_SA_Watanabe}%
  \BibitemOpen
  \bibfield  {author} {\bibinfo {author} {\bibfnamefont {H.}~\bibnamefont {Watanabe}}, \bibinfo {author} {\bibfnamefont {H.~C.}\ \bibnamefont {Po}},\ and\ \bibinfo {author} {\bibfnamefont {A.}~\bibnamefont {Vishwanath}},\ }\bibfield  {title} {\bibinfo {title} {{Structure and topology of band structures in the 1651 magnetic space groups}},\ }\href {http://advances.sciencemag.org/content/4/8/eaat8685} {\bibfield  {journal} {\bibinfo  {journal} {Sci. Adv.}\ }\textbf {\bibinfo {volume} {4}},\ \bibinfo {pages} {eaat8685} (\bibinfo {year} {2018})}\BibitemShut {NoStop}%
\bibitem [{\citenamefont {Bradlyn}\ \emph {et~al.}(2017)\citenamefont {Bradlyn}, \citenamefont {Elcoro}, \citenamefont {Cano}, \citenamefont {Vergniory}, \citenamefont {Wang}, \citenamefont {Felser}, \citenamefont {Aroyo},\ and\ \citenamefont {Bernevig}}]{Bradlyn_2017}%
  \BibitemOpen
  \bibfield  {author} {\bibinfo {author} {\bibfnamefont {B.}~\bibnamefont {Bradlyn}}, \bibinfo {author} {\bibfnamefont {L.}~\bibnamefont {Elcoro}}, \bibinfo {author} {\bibfnamefont {J.}~\bibnamefont {Cano}}, \bibinfo {author} {\bibfnamefont {M.~G.}\ \bibnamefont {Vergniory}}, \bibinfo {author} {\bibfnamefont {Z.}~\bibnamefont {Wang}}, \bibinfo {author} {\bibfnamefont {C.}~\bibnamefont {Felser}}, \bibinfo {author} {\bibfnamefont {M.~I.}\ \bibnamefont {Aroyo}},\ and\ \bibinfo {author} {\bibfnamefont {B.~A.}\ \bibnamefont {Bernevig}},\ }\bibfield  {title} {\bibinfo {title} {Topological quantum chemistry},\ }\href {https://doi.org/10.1038/nature23268} {\bibfield  {journal} {\bibinfo  {journal} {Nature}\ }\textbf {\bibinfo {volume} {547}},\ \bibinfo {pages} {298} (\bibinfo {year} {2017})}\BibitemShut {NoStop}%
\bibitem [{\citenamefont {Cano}\ \emph {et~al.}(2018)\citenamefont {Cano}, \citenamefont {Bradlyn}, \citenamefont {Wang}, \citenamefont {Elcoro}, \citenamefont {Vergniory}, \citenamefont {Felser}, \citenamefont {Aroyo},\ and\ \citenamefont {Bernevig}}]{PhysRevB.97.035139}%
  \BibitemOpen
  \bibfield  {author} {\bibinfo {author} {\bibfnamefont {J.}~\bibnamefont {Cano}}, \bibinfo {author} {\bibfnamefont {B.}~\bibnamefont {Bradlyn}}, \bibinfo {author} {\bibfnamefont {Z.}~\bibnamefont {Wang}}, \bibinfo {author} {\bibfnamefont {L.}~\bibnamefont {Elcoro}}, \bibinfo {author} {\bibfnamefont {M.~G.}\ \bibnamefont {Vergniory}}, \bibinfo {author} {\bibfnamefont {C.}~\bibnamefont {Felser}}, \bibinfo {author} {\bibfnamefont {M.~I.}\ \bibnamefont {Aroyo}},\ and\ \bibinfo {author} {\bibfnamefont {B.~A.}\ \bibnamefont {Bernevig}},\ }\bibfield  {title} {\bibinfo {title} {Building blocks of topological quantum chemistry: Elementary band representations},\ }\href {https://doi.org/10.1103/PhysRevB.97.035139} {\bibfield  {journal} {\bibinfo  {journal} {Phys. Rev. B}\ }\textbf {\bibinfo {volume} {97}},\ \bibinfo {pages} {035139} (\bibinfo {year} {2018})}\BibitemShut {NoStop}%
\bibitem [{\citenamefont {Aroyo}\ \emph {et~al.}(2006)\citenamefont {Aroyo}, \citenamefont {Kirov}, \citenamefont {Capillas}, \citenamefont {Perez-Mato},\ and\ \citenamefont {Wondratschek}}]{Aroyo:xo5013}%
  \BibitemOpen
  \bibfield  {author} {\bibinfo {author} {\bibfnamefont {M.~I.}\ \bibnamefont {Aroyo}}, \bibinfo {author} {\bibfnamefont {A.}~\bibnamefont {Kirov}}, \bibinfo {author} {\bibfnamefont {C.}~\bibnamefont {Capillas}}, \bibinfo {author} {\bibfnamefont {J.~M.}\ \bibnamefont {Perez-Mato}},\ and\ \bibinfo {author} {\bibfnamefont {H.}~\bibnamefont {Wondratschek}},\ }\bibfield  {title} {\bibinfo {title} {{Bilbao Crystallographic Server. II. Representations of crystallographic point groups and space groups}},\ }\href {https://doi.org/10.1107/S0108767305040286} {\bibfield  {journal} {\bibinfo  {journal} {Acta Crystallographica Section A}\ }\textbf {\bibinfo {volume} {62}},\ \bibinfo {pages} {115} (\bibinfo {year} {2006})}\BibitemShut {NoStop}%
\bibitem [{\citenamefont {Elcoro}\ \emph {et~al.}(2021)\citenamefont {Elcoro}, \citenamefont {Wieder}, \citenamefont {Song}, \citenamefont {Xu}, \citenamefont {Bradlyn},\ and\ \citenamefont {Bernevig}}]{MTQC}%
  \BibitemOpen
  \bibfield  {author} {\bibinfo {author} {\bibfnamefont {L.}~\bibnamefont {Elcoro}}, \bibinfo {author} {\bibfnamefont {B.~J.}\ \bibnamefont {Wieder}}, \bibinfo {author} {\bibfnamefont {Z.}~\bibnamefont {Song}}, \bibinfo {author} {\bibfnamefont {Y.}~\bibnamefont {Xu}}, \bibinfo {author} {\bibfnamefont {B.}~\bibnamefont {Bradlyn}},\ and\ \bibinfo {author} {\bibfnamefont {B.~A.}\ \bibnamefont {Bernevig}},\ }\bibfield  {title} {\bibinfo {title} {Magnetic topological quantum chemistry},\ }\href {https://doi.org/10.1038/s41467-021-26241-8} {\bibfield  {journal} {\bibinfo  {journal} {Nature Communications}\ }\textbf {\bibinfo {volume} {12}},\ \bibinfo {pages} {5965} (\bibinfo {year} {2021})}\BibitemShut {NoStop}%
\bibitem [{\citenamefont {Tang}\ \emph {et~al.}(2019{\natexlab{a}})\citenamefont {Tang}, \citenamefont {Po}, \citenamefont {Vishwanath},\ and\ \citenamefont {Wan}}]{catalogue0}%
  \BibitemOpen
  \bibfield  {author} {\bibinfo {author} {\bibfnamefont {F.}~\bibnamefont {Tang}}, \bibinfo {author} {\bibfnamefont {H.~C.}\ \bibnamefont {Po}}, \bibinfo {author} {\bibfnamefont {A.}~\bibnamefont {Vishwanath}},\ and\ \bibinfo {author} {\bibfnamefont {X.}~\bibnamefont {Wan}},\ }\bibfield  {title} {\bibinfo {title} {{Efficient topological materials discovery using symmetry indicators}},\ }\href {https://doi.org/10.1038/s41567-019-0418-7} {\bibfield  {journal} {\bibinfo  {journal} {Nat. Phys.}\ }\textbf {\bibinfo {volume} {15}},\ \bibinfo {pages} {470} (\bibinfo {year} {2019}{\natexlab{a}})}\BibitemShut {NoStop}%
\bibitem [{\citenamefont {Zhang}\ \emph {et~al.}(2019)\citenamefont {Zhang}, \citenamefont {Jiang}, \citenamefont {Song}, \citenamefont {Huang}, \citenamefont {He}, \citenamefont {Fang}, \citenamefont {Weng},\ and\ \citenamefont {Fang}}]{catalogue1}%
  \BibitemOpen
  \bibfield  {author} {\bibinfo {author} {\bibfnamefont {T.}~\bibnamefont {Zhang}}, \bibinfo {author} {\bibfnamefont {Y.}~\bibnamefont {Jiang}}, \bibinfo {author} {\bibfnamefont {Z.}~\bibnamefont {Song}}, \bibinfo {author} {\bibfnamefont {H.}~\bibnamefont {Huang}}, \bibinfo {author} {\bibfnamefont {Y.}~\bibnamefont {He}}, \bibinfo {author} {\bibfnamefont {Z.}~\bibnamefont {Fang}}, \bibinfo {author} {\bibfnamefont {H.}~\bibnamefont {Weng}},\ and\ \bibinfo {author} {\bibfnamefont {C.}~\bibnamefont {Fang}},\ }\bibfield  {title} {\bibinfo {title} {{Catalogue of topological electronic materials}},\ }\href {https://doi.org/10.1038/s41586-019-0944-6} {\bibfield  {journal} {\bibinfo  {journal} {Nature}\ }\textbf {\bibinfo {volume} {566}},\ \bibinfo {pages} {475} (\bibinfo {year} {2019})}\BibitemShut {NoStop}%
\bibitem [{\citenamefont {Tang}\ \emph {et~al.}(2019{\natexlab{b}})\citenamefont {Tang}, \citenamefont {Po}, \citenamefont {Vishwanath},\ and\ \citenamefont {Wan}}]{catalogue2}%
  \BibitemOpen
  \bibfield  {author} {\bibinfo {author} {\bibfnamefont {F.}~\bibnamefont {Tang}}, \bibinfo {author} {\bibfnamefont {H.~C.}\ \bibnamefont {Po}}, \bibinfo {author} {\bibfnamefont {A.}~\bibnamefont {Vishwanath}},\ and\ \bibinfo {author} {\bibfnamefont {X.}~\bibnamefont {Wan}},\ }\bibfield  {title} {\bibinfo {title} {{Comprehensive search for topological materials using symmetry indicators}},\ }\href {https://doi.org/10.1038/s41586-019-0937-5} {\bibfield  {journal} {\bibinfo  {journal} {Nature}\ }\textbf {\bibinfo {volume} {566}},\ \bibinfo {pages} {486} (\bibinfo {year} {2019}{\natexlab{b}})}\BibitemShut {NoStop}%
\bibitem [{\citenamefont {Vergniory}\ \emph {et~al.}(2019)\citenamefont {Vergniory}, \citenamefont {Elcoro}, \citenamefont {Felser}, \citenamefont {Regnault}, \citenamefont {Bernevig},\ and\ \citenamefont {Wang}}]{catalogue3}%
  \BibitemOpen
  \bibfield  {author} {\bibinfo {author} {\bibfnamefont {M.~G.}\ \bibnamefont {Vergniory}}, \bibinfo {author} {\bibfnamefont {L.}~\bibnamefont {Elcoro}}, \bibinfo {author} {\bibfnamefont {C.}~\bibnamefont {Felser}}, \bibinfo {author} {\bibfnamefont {N.}~\bibnamefont {Regnault}}, \bibinfo {author} {\bibfnamefont {B.~A.}\ \bibnamefont {Bernevig}},\ and\ \bibinfo {author} {\bibfnamefont {Z.}~\bibnamefont {Wang}},\ }\bibfield  {title} {\bibinfo {title} {{A complete catalogue of high-quality topological materials}},\ }\href {https://doi.org/10.1038/s41586-019-0954-4} {\bibfield  {journal} {\bibinfo  {journal} {Nature}\ }\textbf {\bibinfo {volume} {566}},\ \bibinfo {pages} {480} (\bibinfo {year} {2019})}\BibitemShut {NoStop}%
\bibitem [{\citenamefont {Tang}\ \emph {et~al.}(2019{\natexlab{c}})\citenamefont {Tang}, \citenamefont {Po}, \citenamefont {Vishwanath},\ and\ \citenamefont {Wan}}]{catalogue4}%
  \BibitemOpen
  \bibfield  {author} {\bibinfo {author} {\bibfnamefont {F.}~\bibnamefont {Tang}}, \bibinfo {author} {\bibfnamefont {H.~C.}\ \bibnamefont {Po}}, \bibinfo {author} {\bibfnamefont {A.}~\bibnamefont {Vishwanath}},\ and\ \bibinfo {author} {\bibfnamefont {X.}~\bibnamefont {Wan}},\ }\bibfield  {title} {\bibinfo {title} {{Topological materials discovery by large-order symmetry indicators}},\ }\href {https://advances.sciencemag.org/content/5/3/eaau8725} {\bibfield  {journal} {\bibinfo  {journal} {Sci. Adv.}\ }\textbf {\bibinfo {volume} {5}},\ \bibinfo {pages} {eaau8725} (\bibinfo {year} {2019}{\natexlab{c}})}\BibitemShut {NoStop}%
\bibitem [{\citenamefont {Xu}\ \emph {et~al.}(2020)\citenamefont {Xu}, \citenamefont {Elcoro}, \citenamefont {Song}, \citenamefont {Wieder}, \citenamefont {Vergniory}, \citenamefont {Regnault}, \citenamefont {Chen}, \citenamefont {Felser},\ and\ \citenamefont {Bernevig}}]{catalogue6}%
  \BibitemOpen
  \bibfield  {author} {\bibinfo {author} {\bibfnamefont {Y.}~\bibnamefont {Xu}}, \bibinfo {author} {\bibfnamefont {L.}~\bibnamefont {Elcoro}}, \bibinfo {author} {\bibfnamefont {Z.-D.}\ \bibnamefont {Song}}, \bibinfo {author} {\bibfnamefont {B.~J.}\ \bibnamefont {Wieder}}, \bibinfo {author} {\bibfnamefont {M.~G.}\ \bibnamefont {Vergniory}}, \bibinfo {author} {\bibfnamefont {N.}~\bibnamefont {Regnault}}, \bibinfo {author} {\bibfnamefont {Y.}~\bibnamefont {Chen}}, \bibinfo {author} {\bibfnamefont {C.}~\bibnamefont {Felser}},\ and\ \bibinfo {author} {\bibfnamefont {B.~A.}\ \bibnamefont {Bernevig}},\ }\bibfield  {title} {\bibinfo {title} {High-throughput calculations of magnetic topological materials},\ }\href {https://doi.org/10.1038/s41586-020-2837-0} {\bibfield  {journal} {\bibinfo  {journal} {Nature}\ }\textbf {\bibinfo {volume} {586}},\ \bibinfo {pages} {702} (\bibinfo {year} {2020})}\BibitemShut {NoStop}%
\bibitem [{\citenamefont {Fang}\ and\ \citenamefont {Cano}(2021)}]{PhysRevB.103.165109}%
  \BibitemOpen
  \bibfield  {author} {\bibinfo {author} {\bibfnamefont {Y.}~\bibnamefont {Fang}}\ and\ \bibinfo {author} {\bibfnamefont {J.}~\bibnamefont {Cano}},\ }\bibfield  {title} {\bibinfo {title} {Filling anomaly for general two- and three-dimensional ${C}_{4}$ symmetric lattices},\ }\href {https://doi.org/10.1103/PhysRevB.103.165109} {\bibfield  {journal} {\bibinfo  {journal} {Phys. Rev. B}\ }\textbf {\bibinfo {volume} {103}},\ \bibinfo {pages} {165109} (\bibinfo {year} {2021})}\BibitemShut {NoStop}%
\bibitem [{\citenamefont {Bulatov}(2002)}]{bridges2002:320}%
  \BibitemOpen
  \bibfield  {author} {\bibinfo {author} {\bibfnamefont {V.}~\bibnamefont {Bulatov}},\ }\bibfield  {title} {\bibinfo {title} {About enumeration of isogonal polyhedral families - abstract},\ }in\ \href {http://archive.bridgesmathart.org/2002/bridges2002-320.html} {\emph {\bibinfo {booktitle} {Bridges: Mathematical Connections in Art, Music, and Science}}},\ \bibinfo {editor} {edited by\ \bibinfo {editor} {\bibfnamefont {R.}~\bibnamefont {Sarhangi}}}\ (\bibinfo  {publisher} {Bridges Conference},\ \bibinfo {address} {Southwestern College, Winfield, Kansas},\ \bibinfo {year} {2002})\ pp.\ \bibinfo {pages} {320--320}\BibitemShut {NoStop}%
\bibitem [{\citenamefont {Robertson}\ \emph {et~al.}(1970)\citenamefont {Robertson}, \citenamefont {Carter},\ and\ \citenamefont {Morton}}]{ROBERTSON197079}%
  \BibitemOpen
  \bibfield  {author} {\bibinfo {author} {\bibfnamefont {S.}~\bibnamefont {Robertson}}, \bibinfo {author} {\bibfnamefont {S.}~\bibnamefont {Carter}},\ and\ \bibinfo {author} {\bibfnamefont {H.}~\bibnamefont {Morton}},\ }\bibfield  {title} {\bibinfo {title} {Finite orthogonal symmetry},\ }\href {https://doi.org/https://doi.org/10.1016/0040-9383(70)90052-2} {\bibfield  {journal} {\bibinfo  {journal} {Topology}\ }\textbf {\bibinfo {volume} {9}},\ \bibinfo {pages} {79} (\bibinfo {year} {1970})}\BibitemShut {NoStop}%
\bibitem [{\citenamefont {Cromwell}(1997)}]{polyhedra}%
  \BibitemOpen
  \bibfield  {author} {\bibinfo {author} {\bibfnamefont {P.~R.}\ \bibnamefont {Cromwell}},\ }\href@noop {} {\emph {\bibinfo {title} {Polyhedra}}}\ (\bibinfo  {publisher} {Cambridge University Press, Cambridge},\ \bibinfo {year} {1997})\BibitemShut {NoStop}%
\bibitem [{\citenamefont {Robertson}\ and\ \citenamefont {Carter}(1970)}]{https://doi.org/10.1112/jlms/s2-2.1.125}%
  \BibitemOpen
  \bibfield  {author} {\bibinfo {author} {\bibfnamefont {S.~A.}\ \bibnamefont {Robertson}}\ and\ \bibinfo {author} {\bibfnamefont {S.}~\bibnamefont {Carter}},\ }\bibfield  {title} {\bibinfo {title} {On the platonic and archimedean solids},\ }\href {https://doi.org/https://doi.org/10.1112/jlms/s2-2.1.125} {\bibfield  {journal} {\bibinfo  {journal} {Journal of the London Mathematical Society}\ }\textbf {\bibinfo {volume} {s2-2}},\ \bibinfo {pages} {125} (\bibinfo {year} {1970})}\BibitemShut {NoStop}%
\bibitem [{\citenamefont {Grünbaum}(1997)}]{IsogonalPrismatoids}%
  \BibitemOpen
  \bibfield  {author} {\bibinfo {author} {\bibfnamefont {B.}~\bibnamefont {Grünbaum}},\ }\href {https://doi.org/10.1007/PL00009307} {\bibinfo {title} {Isogonal prismatoids, {D}iscrete {C}omput. {G}eom. 18, 13-52}} (\bibinfo {year} {1997})\BibitemShut {NoStop}%
\bibitem [{\citenamefont {Leopold}(2017)}]{Leopold}%
  \BibitemOpen
  \bibfield  {author} {\bibinfo {author} {\bibfnamefont {U.}~\bibnamefont {Leopold}},\ }\bibfield  {title} {\bibinfo {title} {Vertex-transitive polyhedra of higher genus, i},\ }\href {https://doi.org/10.1007/s00454-016-9828-9} {\bibfield  {journal} {\bibinfo  {journal} {Discrete Comput. Geom.}\ }\textbf {\bibinfo {volume} {57}},\ \bibinfo {pages} {125} (\bibinfo {year} {2017})}\BibitemShut {NoStop}%
\bibitem [{\citenamefont {Gr{\"u}nbaum}\ and\ \citenamefont {Shephard}(1984)}]{Grunbaum}%
  \BibitemOpen
  \bibfield  {author} {\bibinfo {author} {\bibfnamefont {B.}~\bibnamefont {Gr{\"u}nbaum}}\ and\ \bibinfo {author} {\bibfnamefont {G.}~\bibnamefont {Shephard}},\ }\bibfield  {title} {\bibinfo {title} {Polyhedra with transitivity properties},\ }\href@noop {} {\bibfield  {journal} {\bibinfo  {journal} {C. R. Math. Rep. Acad. Sci. Canada}\ }\textbf {\bibinfo {volume} {6}},\ \bibinfo {pages} {61–66} (\bibinfo {year} {1984})}\BibitemShut {NoStop}%
\bibitem [{\citenamefont {Ono}\ and\ \citenamefont {Shiozaki}(2023)}]{Ono=Shiozaki_top_inv}%
  \BibitemOpen
  \bibfield  {author} {\bibinfo {author} {\bibfnamefont {S.}~\bibnamefont {Ono}}\ and\ \bibinfo {author} {\bibfnamefont {K.}~\bibnamefont {Shiozaki}},\ }\href@noop {} {\bibinfo {title} {{Towards complete characterization of topological insulators and superconductors: A systematic construction of topological invariants based on Atiyah-Hirzebruch spectral sequence}}} (\bibinfo {year} {2023}),\ \Eprint {https://arxiv.org/abs/2311.15814} {arXiv:2311.15814 [cond-mat.mes-hall]} \BibitemShut {NoStop}%
\bibitem [{\citenamefont {Vanderbilt}\ and\ \citenamefont {King-Smith}(1993)}]{PhysRevB.48.4442}%
  \BibitemOpen
  \bibfield  {author} {\bibinfo {author} {\bibfnamefont {D.}~\bibnamefont {Vanderbilt}}\ and\ \bibinfo {author} {\bibfnamefont {R.~D.}\ \bibnamefont {King-Smith}},\ }\bibfield  {title} {\bibinfo {title} {Electric polarization as a bulk quantity and its relation to surface charge},\ }\href {https://doi.org/10.1103/PhysRevB.48.4442} {\bibfield  {journal} {\bibinfo  {journal} {Phys. Rev. B}\ }\textbf {\bibinfo {volume} {48}},\ \bibinfo {pages} {4442} (\bibinfo {year} {1993})}\BibitemShut {NoStop}%
\bibitem [{\citenamefont {King-Smith}\ and\ \citenamefont {Vanderbilt}(1993)}]{PhysRevB.47.1651}%
  \BibitemOpen
  \bibfield  {author} {\bibinfo {author} {\bibfnamefont {R.~D.}\ \bibnamefont {King-Smith}}\ and\ \bibinfo {author} {\bibfnamefont {D.}~\bibnamefont {Vanderbilt}},\ }\bibfield  {title} {\bibinfo {title} {Theory of polarization of crystalline solids},\ }\href {https://doi.org/10.1103/PhysRevB.47.1651} {\bibfield  {journal} {\bibinfo  {journal} {Phys. Rev. B}\ }\textbf {\bibinfo {volume} {47}},\ \bibinfo {pages} {1651} (\bibinfo {year} {1993})}\BibitemShut {NoStop}%
\bibitem [{\citenamefont {Kresse}\ and\ \citenamefont {Furthm{\"u}ller}(1996{\natexlab{a}})}]{kresse1996efficient}%
  \BibitemOpen
  \bibfield  {author} {\bibinfo {author} {\bibfnamefont {G.}~\bibnamefont {Kresse}}\ and\ \bibinfo {author} {\bibfnamefont {J.}~\bibnamefont {Furthm{\"u}ller}},\ }\bibfield  {title} {\bibinfo {title} {Efficient iterative schemes for $ab$ $initio$ total-energy calculations using a plane-wave basis set},\ }\href {https://doi.org/10.1103/PhysRevB.54.11169} {\bibfield  {journal} {\bibinfo  {journal} {Phys. Rev. B}\ }\textbf {\bibinfo {volume} {54}},\ \bibinfo {pages} {11169} (\bibinfo {year} {1996}{\natexlab{a}})}\BibitemShut {NoStop}%
\bibitem [{\citenamefont {Kresse}\ and\ \citenamefont {Hafner}(1994)}]{kresse1994ab}%
  \BibitemOpen
  \bibfield  {author} {\bibinfo {author} {\bibfnamefont {G.}~\bibnamefont {Kresse}}\ and\ \bibinfo {author} {\bibfnamefont {J.}~\bibnamefont {Hafner}},\ }\bibfield  {title} {\bibinfo {title} {$ab$ $initio$ molecular-dynamics simulation of the liquid-metal--amorphous-semiconductor transition in germanium},\ }\href {https://doi.org/10.1103/PhysRevB.49.14251} {\bibfield  {journal} {\bibinfo  {journal} {Phys. Rev. B}\ }\textbf {\bibinfo {volume} {49}},\ \bibinfo {pages} {14251} (\bibinfo {year} {1994})}\BibitemShut {NoStop}%
\bibitem [{\citenamefont {Kresse}\ and\ \citenamefont {Furthm{\"u}ller}(1996{\natexlab{b}})}]{kresse1996efficiency}%
  \BibitemOpen
  \bibfield  {author} {\bibinfo {author} {\bibfnamefont {G.}~\bibnamefont {Kresse}}\ and\ \bibinfo {author} {\bibfnamefont {J.}~\bibnamefont {Furthm{\"u}ller}},\ }\bibfield  {title} {\bibinfo {title} {Efficiency of $ab-initio$ total energy calculations for metals and semiconductors using a plane-wave basis set},\ }\href {https://doi.org/https://doi.org/10.1016/0927-0256(96)00008-0} {\bibfield  {journal} {\bibinfo  {journal} {Comput. Mater. Sci.}\ }\textbf {\bibinfo {volume} {6}},\ \bibinfo {pages} {15} (\bibinfo {year} {1996}{\natexlab{b}})}\BibitemShut {NoStop}%
\bibitem [{\citenamefont {Kresse}\ and\ \citenamefont {Hafner}(1993)}]{kresse1993ab}%
  \BibitemOpen
  \bibfield  {author} {\bibinfo {author} {\bibfnamefont {G.}~\bibnamefont {Kresse}}\ and\ \bibinfo {author} {\bibfnamefont {J.}~\bibnamefont {Hafner}},\ }\bibfield  {title} {\bibinfo {title} {$ab$ $initio$ molecular dynamics for liquid metals},\ }\href {https://doi.org/10.1103/PhysRevB.47.558} {\bibfield  {journal} {\bibinfo  {journal} {Phy. Rev. B}\ }\textbf {\bibinfo {volume} {47}},\ \bibinfo {pages} {558} (\bibinfo {year} {1993})}\BibitemShut {NoStop}%
\end{thebibliography}%
\end{document}